# MacWilliams type identities for some new $m$-spotty weight enumerators over finite commutative Frobenius rings[*]

Minjia Shi

*School of Mathematical Sciences of Anhui University, 230601 Hefei, Anhui, China*

**Abstract** Past few years have seen an extensive use of RAM chips with wide I/O data (e.g. 16, 32, 64 bits) in computer memory systems. These chips are highly vulnerable to a special type of byte error, called an $m$-spotty byte error, which can be effectively detected or corrected using byte error-control codes. The MacWilliams identity provides the relationship between the weight distribution of a code and that of its dual. This paper introduces $m$-spotty Hamming weight enumerator, joint $m$-spotty Hamming weight enumerator and split $m$-spotty Hamming weight enumerator for byte error-control codes over finite commutative Frobenius rings as well as $m$-spotty Lee weight enumerator over an infinite family of rings. In addition, MacWilliams type identities are also derived for these enumerators.

## 1 Introduction

The error-control codes have a significant role in improving reliability in communications and computer memory system [4]. For the past few years, there has been an increased usage of high-density RAM chips with wide I/O data, called a byte, in computer memory systems. These chips are highly vulnerable to multiple random bit errors when exposed to strong electromagnetic waves, radio-active particles or high-energy cosmic rays. To overcome this, a new type of byte error known as spotty byte error has been introduced in which the error occurs at random $t$ or fewer bits within a $b$-bit byte [16], if more than one spotty byte error occur within a $b$-bit byte, then it is known as a multiple spotty byte error or $m$-spotty byte error [15]. To determine the error-detecting and error-correcting capabilities of a code, some special types of polynomials, called weight enumerators, are studied.

One of the most celebrated results in the coding theory is the MacWilliams identity that describes how the weight enumerator of a linear code and the weight enumerator of the dual code relate to each other. This identity has found widespread application in the coding theory [5], especially in the study of self-dual codes [2]. Recently various weight enumerators with respect to $m$-spotty Hamming (Lee) weight have been introduced and studied. Suzuki et al.[15] defined Hamming weight enumerator for binary byte error-control codes, and proved a MacWilliams identity for it. Özen and Siap [6] and Siap [13] extended this result to arbitrary finite fields and to the ring $\mathbb{F}_2+u\mathbb{F}_2$ with $u^2 = 0$, respectively, the later results were generalized to $\mathbb{F}_2+u\mathbb{F}_2+\cdots+u^{m-1}\mathbb{F}_2$ with $u^m = 0$ by Shi [12]. Siap [14] defined $m$-spotty Lee weight and $m$-spotty Lee weight enumerator of byte error-control codes over $\mathbb{Z}_4$ and derived a MacWilliams identity. Sharma et al. introduced joint $m$-spotty weight enumerators of two byte error-control codes over the ring of integers modulo $\ell$ and over arbitrary finite fields with respect to $m$-spotty Hamming weight [8], $m$-spotty Lee weight [9] and $r$-fold joint $m$-spotty weight [10]. Özen and Siap [7] proved a MacWilliams identity for the $m$-spotty RT weight enumerators of binary codes, which was generalized to the case of finite commutative Frobenius rings by Shi [11]. But MaWilliams type identities for $m$-spotty Hamming (Lee) weight enumerators over finite commutative Frobenius ring has not been considered to the best of our knowledge.

Throughout this paper, we let ring $R_k$ be a finite commutative Frobenius ring or an infinite family of rings (see Example 2.4). The organization of this paper is as follows: In Section 2, we state some preliminaries which we need to prove our main results. Section 3 presents a MacWilliams identity for $m$-spotty Hamming weight enumerator over $R_k$. Section 4 and Section 5 determine MacWilliams identities for joint $m$-spotty weight enumerator and split $m$-spotty Hamming weight enumerator over $R_k$, respectively. Finally, we present a MacWilliams identity for $m$-spotty Lee weight enumerator over $R_k$ in Section 6. We also illustrate our results with some examples.

---

[*]**The paper was submitted for reviewing on 30th March**.
This research is supported by NNSF of China (61202068, 11126174), Talents youth Fund of Anhui Province Universities (2012SQRL020ZD). The author is with the School of Mathematical Sciences, Anhui University, Anhui 230601, China, Associate Professor. (e-mail: smjwcl.good@163.com).

## 2 Preliminaries

In this section, we begin by giving some basic definitions that we need to derive our results. Let $R_k$ be a finite commutative Frobenius ring with unity and $N$ be a positive integer. Let us recall some basic knowledge about $R_k$ as described in [3]. The finite commutative ring $R_k$ is called a *Frobenius ring* if $R_k$ is self-injective (i.e., the regular module is injective), or equivalently, $(C^\perp)^\perp = C$ for any submodule $C$ of any free $R_k$-module $R_k^n$, where $C^\perp$ denotes the orthogonal submodule of $C$ with respect to the usual Euclidean inner product on $R_k^n$. Moreover, in this case, $|C^\perp||C| = |R_k|^n$ for any submodule $C$ of $R_k^n$, where $|C|$ denotes the cardinality of $C$. This is one of the reasons why only finite Frobenius rings are suitable for coding alphabets. The reader may refer to [17,18] for more details on Frobenius rings.

**Remark 2.1**. Many well-known finite rings are finite Frobenius rings. Here are a number of examples of finite commutative Frobenius rings with their generating characters.

(i) Let $R_k = \mathbb{F}$ be a finite field. A generating character $\chi$ on $R_k$ is given by $\chi(x) = \xi^{Tr(x)}$, where $\xi = e^{\frac{2\pi i}{p}}$ and $\text{Tr}: \mathbb{F}_\ell \to \mathbb{F}_p$ is the trace function from $\mathbb{F}$ to $\mathbb{F}_p$.

(ii) Let $R_k = \mathbb{Z}_\ell$. Set $\xi = e^{\frac{2\pi i}{\ell}}$. Then $\chi(x) = \xi^x$, $x \in \mathbb{Z}_\ell$, is a generating character.

(iii) The finite direct sum of Frobenius rings is Frobenius. If $R_1, \cdots, R_n$ each has generating characters $\chi_1, \cdots, \chi_n$, then $R_k = \oplus R_i$ has generating character $\chi = \prod \chi_i$.

(iv) A Galois ring $R_k = GR(p^n, r) \cong \mathbb{Z}_{p^n}[x]/\langle f \rangle$ is a Galois extension of $\mathbb{Z}_{p^n}$ of degree $r$, where $f$ is a monic irreducible polynomial in $\mathbb{Z}_{p^n}[x]$ of degree $r$. Any element $a$ of $R_k$ is represented by a unique polynomial $r = \sum_{i=1}^{r-1} a_i x^i$, with $a_i \in \mathbb{Z}_{p^n}$. Set $\xi = e^{\frac{2\pi i}{p^n}}$. Then $\chi(a) = \xi^{a_{r-1}}$.

(v) Let $R_k$ be a finite chain ring with maximal ideal $\langle u \rangle$ and let its residue field $R_k/\langle u \rangle$ be $\mathbb{F}_{p^n}$, i.e, $R_k = \mathbb{F}_q + u\mathbb{F}_q + \cdots + u^{k-1}\mathbb{F}_q$. Any element $r$ of $R_k$ is represented by a unique polynomial $r = \sum_{i=1}^{r-1} a_i u^i$, with $r_i \in \mathbb{F}_q$. Set $\xi = e^{\frac{2\pi i}{q}}$. Then $\chi_r = \xi^{a_{r-1}}$.

Let $(G, +)$ be a finite abelian group and $V$ be a vector space over the complex numbers. The set $\widehat{G}$ of all characters of $G$ forms an abelian group under pointwise miltiplication. For any function $f : G \longrightarrow V$, define its *Fourier transform* $\widehat{f} : \widehat{G} \longrightarrow V$ by

$$\widehat{f}(\pi) = \sum_{x \in G} \pi(x) f(x), \pi \in \widehat{G}.$$

Given a subgroup $H \subseteq G$, define an *annihilator* $(\widehat{G} : H) = \{\pi \in \widehat{G} : \pi(H) = 1\}$. Moreover, we have $|(\widehat{G} : H)| = |G|/|H|$.

*The Poission summation formula* relates the sums of a function over a subgroup to the sum of its Fourier transform over the annihilator of the subgroup. The following lemma can be found in [18], which plays an impormant role in deriving the MacWilliams identity for $m$-spotty Hamming weight enumerator.

**Lemma 2.2**. (Poisson Summation Formula) Let $H \subset G$ be a subgroup, and let $f : G \longrightarrow V$ be any function from $G$ to a complex vector space $V$. Then

$$\sum_{x \in H} f(x) = \frac{1}{|(\widehat{G} : H)|} \sum_{\pi \in (\widehat{G}:H)} \widehat{f}(\pi).$$

Hereinafter, codes will be taken to be of length $N$ where $N$ is a multiple of byte length $b$, i.e. $N = nb$. Let $\mathbf{c} = (\mathbf{c}_1, \mathbf{c}_2, \cdots, \mathbf{c}_N)$ and $v = (\mathbf{v}_1, \mathbf{v}_2, \cdots, \mathbf{v}_N)$ be two elements of $R_k^N$. The inner product of $\mathbf{c}$ and $\mathbf{v}$, denoted by $\langle \mathbf{c}, \mathbf{v} \rangle$, is defined as follows:

$$\langle \mathbf{c}, \mathbf{v} \rangle = \sum_{i=1}^{n} \langle c_i, v_i \rangle = \sum_{i=1}^{n} \Big( \sum_{j=1}^{b} c_{i,j} v_{i,j} \Big).$$

Here, $\langle \mathbf{c}_i, \mathbf{v}_i \rangle = \sum_{j=1}^{b} c_{i,j} v_{i,j}$ denotes the inner product of $\mathbf{c}_i$ and $v_i$, respectively.

Let $C$ be a linear code over $R_k^N$. The set $C^\perp = \{\mathbf{v} \in R_k^N | \langle \mathbf{u}, \mathbf{v} \rangle = 0,\ \text{for all } \mathbf{u} \in C\}$ is also a linear code over $R_k$ and it is called the dual code of $C$.

**Remark 2.3**. The following is an example of a ring that is a not a chain ring but a finite commutative Frobenius ring. We will use this ring to exhibit several of the results of the paper.

**Example 2.4**. Let $R_k = \mathbb{F}_2[u_1, u_2, \cdots, u_k]/\langle u_i^2 = 0, u_i u_j = u_j u_i \rangle$. When $k = 1$, $R_1 = \mathbb{F}_2 + u\mathbb{F}_2$ is a principal ideal ring. When $k = 2$, the ring is $R_2 = \mathbb{F}_2 + u\mathbb{F}_2 + v\mathbb{F}_2 + uv\mathbb{F}_2$ is a nonprincipal ideal ring and we will often use $R_2$ in the following examples. More details about $R_k$ can be found in [1]. The ring can also be defined recursively. Let $R_k = R_{k-1}[u_k]/\langle u_k^2 = 0, u_k u_j = u_j u_k \rangle = R_{k-1} + uR_{k-1}$. For any subset $A \subseteq \{1, 2, \cdots, k\}$ we will fix

$$u_A := \prod_{i \in A} u_i$$

with the convention that $u_\phi = 1$. Then any element of $R_k$ can be represented as

$$\sum_{A \subseteq \{1,2,\cdots,k\}} c_A u_A,\ c_A \in \mathbb{F}_2.$$

Since $(c_A)$ can be thought of as a binary vector of length $2^k$. Let $wt(c_A)$ be the Hamming weight of this vector. Then

$$\chi(r_k) = (-1)^{wt(c_A)}.$$

Throughout this paper, we adopt the notations $|R_k| = \ell$ unless otherwise stated, i.e, denote the size of finite commutative Frobenius ring $R_k$ by $\ell$. In Example 2.4, we have $\ell = 2^{2^k}$.

## 3  MacWilliams type identity for $m$-Spotty Hamming enumerator over finite commutative Frobenius rings

In this section, we extend the results in [6] and [13] to the case over finite commutative Frobenius rings. Let us begin with some definitions.

**Definition 3.1** (see [15]). A spotty byte error is defined as $t$ or fewer bits errors within a $b$-bit byte, where $1 \leq t \leq b$. When none of the bits in a byte is in error, we say that no spotty byte error has occurred.

We can define the $m$-spotty weight and the $m$-spotty distance over $R_k$ as follows.

**Definition 3.2**. Let $\mathbf{e} \in R_k^N$ be an error vector and $\mathbf{e}_i \in R_k^b$ be the $i$-th byte of $\mathbf{e}$, where $N = nb$ and $1 \leq i \leq n$. The number of $t/b$-errors in $\mathbf{e}$, denoted by $w_M(\mathbf{e})$, and called $m$-spotty Hamming weight is defined as $w_M(\mathbf{e}) = \sum_{i=1}^{n} \left\lceil \frac{w_M(\mathbf{e}_i)}{t} \right\rceil$, where $\lceil x \rceil$ denotes the ceiling of $x$ for any real number $x$, i.e., the smallest integer number greater than or equal to $x$. If $t = 1$, this weight, defined by $w_M$, is equal to the Hamming weight. In a similar way, we define the $m$-spotty distance of two codewords $\mathbf{u}$ and $\mathbf{v}$ as $d_M = \sum_{i=0}^{n} \left\lceil \frac{d_H(\mathbf{u}_i, \mathbf{v}_i)}{t} \right\rceil$. Further, it is also straightforward to show that this distance is a metric in $R_k^N$.

**Definition 3.3.** Let $\mathbf{v} = (v_1, v_2, \cdots, v_b) \in R_k^b$. Then the support of $\mathbf{v}$ is defined by supp($\mathbf{v}$)= $\{i | v_i \neq 0\}$ and the complement of supp($\mathbf{v}$) is denoted $\overline{\text{supp}(\mathbf{v})}$.

**Definition 3.4.** Let $\mathbf{c} = (c_1, c_2, \cdots, c_b) \in R_k^b$ and define

$$S_p(\mathbf{c}) = \{\mathbf{v} \in R_k^b | \text{supp}(\mathbf{v}) \subseteq \text{supp}(\mathbf{c}) \text{ and } p = |\text{supp}(\mathbf{v})|\} \text{ and}$$
$$\overline{S}_p(\mathbf{c}) = \{\mathbf{v} \in R_k^b | \text{supp}(\mathbf{v}) \subseteq \overline{\text{supp}(\mathbf{c})} \text{ and } p = |\text{supp}(\mathbf{v})|\}.$$

The following lemmas will be needed later when we are ready to prove our main theorem in this section.

**Lemma 3.5.** Let $H \neq 0$ be an ideal of $R_k$, $a \in R_k$. Then we have $\sum\limits_{a \in H} \chi(a) = 0$ and

$$\sum_{r \in R_k} \chi(ar) = \begin{cases} \ell, & \text{if } a = 0, \\ 0, & \text{if } a \neq 0. \end{cases}$$

*Proof.* We can obtain the first assertion readily by using the definition of character $\chi$. If $a = 0$, then clearly $\chi(ar) = 1$ for all $r \in R_k$ and hence the result follows. Otherwise, if $a$ is a unit, then elements $ar$, for all $r \in R_k$, run over all elements of $R_k$, which forms a trivial ideal $R_k$. If $a$ is a zero divisor, then elements $ar$, for all $r \in R_k$, form a proper ideal of $R_k$. Hence, according to the first assertion in this Lemma, if $a \neq 0$, we have $\sum\limits_{r \in R_k} \chi(ar) = 0$. □

**Lemma 3.6.** Let $\mathbf{v} = (v_1, v_2, \cdots, v_b) \in R_k^b$ with $w(\mathbf{c}) = j \neq 0$ and $p \in \{1, 2, \cdots, j\}$. Then

$$\sum_{\substack{0 \leq w(v) \leq p \\ \text{supp}(v) \subseteq \text{supp}(c)}} \chi(\langle \mathbf{c}, \mathbf{v} \rangle) = 0.$$

*Proof.* Let $\{l_1, l_2, \cdots, l_p\} \subseteq \text{supp}(C)$. If we define a map $\varphi : R_k^p \longrightarrow R_k$ such that $\varphi(v_1, v_2, \cdots, v_p) = c_{l_1} v_1 + \cdots + c_{l_p} v_p$. This is a group homomorphism and the image $\text{Im}(\varphi) = H$ is not zero since $w(c) \neq 0$. Further, $H$ is the nonzero subgroup of $R_k$ generated by $\{c_{l_1}, \cdots, c_{l_p}\}$. Thus, by applying the first group isomorphism theorem, $|R_k^p|/|ker(\varphi)| = |H| \neq \{0\}$. Let $|ker(\varphi)| = m$.

$$\sum_{\substack{0 \leq w(v) \leq p \\ \text{supp}(v) \subseteq \text{supp}(c)}} \chi(\langle \mathbf{c}, \mathbf{v} \rangle) = \sum_{(v_{l_1}, \cdots, v_{l_p})} \chi \left( \sum_{i=1}^{p} c_{l_i} v_{l_i} \right) = m \sum_{h \in H} \chi(h) = 0.$$

This proves the Lemma. □

Similar to [12], applying the method of mathematical induction, we have the follwing lemma .

**Lemma 3.7.** Let $\mathbf{c} = (c_1, c_2, \cdots, c_b) \in R_k^b$ and $w(\mathbf{c}) \neq 0$. For all $p$ positive integers, we let $I_p = \{i_1, i_2, \cdots, i_p\} \subseteq \text{supp}(\mathbf{c})$ and $I_0 = \emptyset$. Then we have

$$\sum_{\substack{\mathbf{v} \in R_k^b \\ \text{supp}(\mathbf{v}) = I_p}} \chi(\langle \mathbf{c}, \mathbf{v} \rangle) = (-1)^p.$$

**Lemma 3.8.** Let $\mathbf{c} = (c_1, c_2, \cdots, c_b) \in R_k^b$ and $w(\mathbf{c}) = j \neq 0$. For all $0 \leq p \leq j$, we have

(i) $\sum\limits_{\mathbf{v} \in S_p(c)} \chi(\langle \mathbf{c}, \mathbf{v} \rangle) = (-1)^p \binom{j}{p}$; (ii) $\sum\limits_{\mathbf{v} \in \overline{S}_p(c)} \chi(\langle \mathbf{c}, \mathbf{v} \rangle) = (\ell - 1)^p \binom{b-j}{p}$.

*Proof.* According to Definition 3.4 and Lemma 3.7, we get

$$\sum_{\mathbf{v} \in S_p(\mathbf{c})} \chi(\langle \mathbf{c}, \mathbf{v} \rangle) = \sum_{I_p \subseteq \mathrm{supp}(\mathbf{c})} \sum_{\mathrm{supp}(\mathbf{v})=I_p} \chi(\langle \mathbf{c}, \mathbf{v} \rangle) = \sum_{I_p \subseteq \mathrm{supp}(\mathbf{c})} (-1)^p = (-1)^p \binom{j}{p}.$$

Since $\mathbf{v} \in \overline{S}_p(\mathbf{c})$ with $\mathrm{supp}(\mathbf{v}) \subseteq \overline{\mathrm{supp}(\mathbf{c})}$, we have $\chi(\langle \mathbf{c}, \mathbf{v} \rangle) = 1$. Further, since $p = |\mathrm{supp}(\mathbf{v})|$, there are $\binom{b-j}{p}$ ways of choosing a subset of size $p$ from the complement of support of $c$ of size $p$. For each subset of size $p$, the sum of characters equals to $(\ell - 1)^p$. This proves the result. □

Following Lemma 3.8, we have the following corollary.

**Corollary 3.9.** Let $\mathbf{c} = (c_1, c_2, \cdots, c_b) \in R_k^b$ and $w(\mathbf{c}) = j$, $0 \leq j_1 \leq j$ and $0 \leq j_2 \leq b - j$. If $S_{j_1, j_2}(\mathbf{c}) = \{\mathbf{v} \in R_k^b | j_1 = |\mathrm{supp}(\mathbf{v}) \cap \mathrm{supp}(\mathbf{c})| \text{ and } j_2 = |\mathrm{supp}(\mathbf{v}) \cap \overline{\mathrm{supp}(\mathbf{c})}|\}$, then

$$\sum_{\mathbf{v} \in s_{j_1, j_2}(c)} \chi(\langle \mathbf{c}, \mathbf{v} \rangle) = (-1)^{j_1} (\ell - 1)^{j_2} \binom{j}{j_1} \binom{b-j}{j_2}.$$

**Lemma 3.10.** Let $\mathbf{c} = (c_1, c_2, \cdots, c_b) \in R_k^b$ and $w(\mathbf{c}) = j$. Then

$$\sum_{\mathbf{v} \in R_k^b} \chi(\langle \mathbf{c}, \mathbf{v} \rangle) z^{\lceil w_M(\mathbf{v})/t \rceil} = \sum_{j_1=0}^{j} \sum_{j_2=0}^{b-j} (-1)^{j_1} (\ell - 1)^{j_2} \binom{j}{j_1} \binom{b-j}{j_2} z^{\lceil (j_1+j_2)/t \rceil}.$$

*Proof.* Since the sum $\sum_{\mathbf{v} \in R_k^b} \chi(\langle \mathbf{c}, \mathbf{v} \rangle) z^{\lceil w_M(\mathbf{v})/t \rceil}$ runs over all $\mathbf{v} \in R_k^b$, we can split the sum according to the set $S_{j_1, j_2}$ where $j_1$ and $j_2$ run through all possible cases. Hence the conclusion of this lemma follows from Corollary 3.9. □

Let $\alpha_j = \#\{i : w(c_i) = j, 1 \leq i \leq n\}$. That is, $\alpha_j$ is the number of bytes having Hamming weight $j$, $0 \leq j \leq b$, in a codeword. The summation of $\alpha_0, \alpha_1, \cdots, \alpha_b$ is equal to the code length in bytes, that is $\sum_{j=0}^{b} \alpha_j = n$. The Hamming weight distribution vector $(\alpha_0, \alpha_1, \cdots, \alpha_b)$ is determined uniquely for the codeword $\mathbf{c}$. Then, the $m$-spotty Hamming weight of the codeword $\mathbf{c}$ is expressed as $w_M(\mathbf{c}) = \sum_{j=0}^{b} \lceil j/t \rceil \cdot \alpha_j$. Let $A_{(\alpha_0, \alpha_1, \cdots, \alpha_b)}$ be the number of codewords with Hamming weight distribution vector $(\alpha_0, \alpha_1, \cdots, \alpha_b)$. For example, let $\mathbf{c} = (0, u, 0, v, 0, u+v, 1, u, 1+u+uv, 0, 0, 0, 1+u+v+uv, u, 0)$ be a codeword over $R_2$ as stated in Example 2.4 with byte $b = 3$ and $n = 4$. Then the Hamming weight distribution vector of the codeword is $(\alpha_0, \alpha_1, \alpha_2, \alpha_3) = (1, 1, 2, 1)$. Therefore, $A_{(1,1,2,1)}$ is the number of codewords with Hamming weight distribution vector $(1, 1, 2, 1)$.

We are now ready to define the $m$-spotty Hamming weight enumerator of a byte error control code over $R_k$.

**Definition 3.11.** The Hamming weight enumerator for $m$-spotty byte error control code $C$ is defined as

$$W(z) = \sum_{\mathbf{c} \in C} z^{w_M(\mathbf{c})}.$$

By using the parameter $A_{(\alpha_0, \alpha_1, \cdots, \alpha_b)}$, which denotes the number of codewords with Hamming weight distribution vector $\{\alpha_0, \alpha_1, \cdots, \alpha_b\}$, $W(z)$ can be expressed as follows:

$$W(z) = \sum_{\substack{(\alpha_0, \ldots, \alpha_b) \\ \alpha_0, \ldots, \alpha_b \geq 0 \\ \alpha_0 + \cdots + \alpha_b = n}} A_{(\alpha_0, \ldots, \alpha_b)} \prod_{j=0}^{b} (z^{\lceil j/t \rceil})^{\alpha_j}.$$

**Theorem 3.12**. Let $C$ be a byte error control code over $R_k$. The relation between the $m$-spotty $t/b$-weight enumerators of $C$ and its dual is given by

$$W^\perp(z) = \sum_{\substack{(\alpha_0,\ldots,\alpha_b) \\ \alpha_0,\ldots,\alpha_b \geq 0 \\ \alpha_0+\cdots+\alpha_b=n}} A^\perp_{(\alpha_0,\ldots,\alpha_b)} \prod_{j=0}^{b} (z^{\lceil j/t \rceil})^{\alpha_j} = \frac{1}{|C|} \sum_{\substack{(\alpha_0,\ldots,\alpha_b) \\ \alpha_0,\ldots,\alpha_b \geq 0 \\ \alpha_0+\cdots+\alpha_b=n}} A_{(\alpha_0,\ldots,\alpha_b)} \prod_{j=0}^{b} (\vartheta_j^{(b,\ell)}(z))^{\alpha_j},$$

where $\vartheta_j^{(b,\ell)}(z) = \sum_{j_1=0}^{j} \sum_{j_2=0}^{b-j} (-1)^{j_1} (\ell-1)^{j_2} \binom{j}{j_1} \binom{b-j}{j_2} z^{\lceil (j_1+j_2)/t \rceil}$.

*Proof.* Given a linear code $C \subset R_k^n$, we apply the Poisson Summation Formula with $G = R_k^n$, $H = C$, and $V = \mathbb{C}[z]$, the polynomial ring over $\mathbb{C}$ in one indeterminate. The first task is to identify the character-theoretic annihilator $(\widehat{G} : H) = (\widehat{R_k}^n : C)$ with $C^\perp$. Let $\rho$ be a generating character of $R_k$. We use $\rho$ to define a homomorphism $\beta : R_k \longrightarrow \widehat{R_k}$. For $r \in R_k$, the character $\beta(r) \in \widehat{R_k}$ has the form $\beta(r)(s) = (r\rho)(s) = \rho(sr)$ for $s \in R_k$. One can verify that $\beta$ is an isomorphism of $R_k$-modules. In particular, wt$(r) =$ wt$(\beta r)$, where wt$(r) = 0$ for $r = 0$, and wt$(r) \neq 0$ for $r \neq 0$.

Extend $\beta$ to an isomorphism $\beta : R_k^n \longrightarrow \widehat{R_k}^n$ of $R_k$-modules, via $\beta(x)(y) = \rho(yx)$, for $x, y \in R_k^n$. Again, wt$(x) =$ wt$(\beta x)$. For $x \in R_k^n$, $\beta(x) \in (\widehat{R_k} : C)$ means $\beta(x)(C) = \beta(C \cdot x) = 1$. This means that the ideal $C \cdot x$ of $R_k$ is contained in ker$(\rho)$. Because $\rho$ is a generating character, which implies that $C \cdot x = 0$. Thus $x \in C^\perp$. The converse is obvious. Thus $C^\perp$ corresponds to $(\widehat{R_k} : C)$ under the isomorphism $\beta$.

Remember that $\beta : R_k^n \longrightarrow \widehat{R_k}^n$ is an isomorphism of $R_k$-modules and $(C^\perp)^\perp = C$. Thus the Poisson Summation Formula becomes

$$\sum_{v \in C^\perp} f(\mathbf{v}) = \frac{1}{|C|} \sum_{\mathbf{c} \in C} \widehat{f}(\mathbf{c}), \tag{1}$$

where the Fourier transform is

$$\widehat{f}(\mathbf{c}) = \sum_{v \in R_k^N} \chi_{\mathbf{c}}(\mathbf{v}) f(\mathbf{v}). \tag{2}$$

Let $f(\mathbf{v}) = z^{w_M(\mathbf{v})}$. Then the function $\widehat{f}(\mathbf{c})$ is calculated as follows:

$$\widehat{f}(\mathbf{c}) = \sum_{\mathbf{v} \in R_k^{nb}} \chi(\langle \mathbf{c}, \mathbf{v} \rangle) z^{w_M(\mathbf{v})}$$

$$= \sum_{\mathbf{v}_1 \in R_k^b} \sum_{\mathbf{v}_2 \in R_k^b} \cdots \sum_{\mathbf{v}_n \in R_k^b} \chi(\langle \mathbf{c}_1, \mathbf{v}_1 \rangle) \chi(\langle \mathbf{c}_2, \mathbf{v}_2 \rangle) \cdots \chi(\langle \mathbf{c}_n, \mathbf{v}_n \rangle) \prod_{i=1}^{n} z^{\lceil w_H(\mathbf{v}_i)/t \rceil}$$

$$= \prod_{i=1}^{n} \left( \sum_{\mathbf{v}_i \in R_k^b} \chi(\langle \mathbf{c}_i, \mathbf{v}_i \rangle) z^{\lceil w_H(\mathbf{v}_i)/t \rceil} \right).$$

By applying Corollary 3.9, we have,

$$\widehat{f}(\mathbf{c}) = \prod_{i=1}^{n} \left( \sum_{j_1=0}^{k_i} \sum_{j_2=0}^{b-k_i} (-1)^{j_1} (\ell-1)^{j_2} \binom{k_i}{j_1} \binom{b-k_i}{j_2} z^{\lceil (j_1+j_2)/t \rceil} \right),$$

where $k_i = w(\mathbf{c}_i)$. Thus

$$\widehat{f}(\mathbf{c}) = \prod_{j=0}^{b} \left( \sum_{j_1=0}^{j} \sum_{j_2=0}^{b-j} (-1)^{j_1} (\ell-1)^{j_2} \binom{j}{j_1} \binom{b-j}{j_2} z^{\lceil (j_1+j_2)/t \rceil} \right)^{\alpha_j(\mathbf{c})},$$

where $\alpha_j(\mathbf{c}) = |\{i | w(\mathbf{c}_i) = j\}|$.

$$\widehat{f}(\mathbf{c}) = \prod_{i=1}^{n} \left( \sum_{j_1=0}^{j} \sum_{j_2=0}^{b-j} (-1)^{j_1} (\ell-1)^{j_2} \binom{j}{j_1} \binom{b-j}{j_2} z^{\lceil (j_1+j_2)/t \rceil} \right)^{\alpha_j(\mathbf{c})}.$$

After rearranging the summations on both sides according to the weight distribution vectors of codewords in $C^\perp$ and $C$ respectively, we have the result

$$W^\perp(z) = \sum_{\substack{(\alpha_0,\ldots,\alpha_b) \\ \alpha_0,\ldots,\alpha_b \geq 0 \\ \alpha_0+\cdots+\alpha_b=n}} A^\perp_{(\alpha_0,\ldots,\alpha_b)} \prod_{j=0}^{b} (z^{\lceil j/t \rceil})^{\alpha_j} = \frac{1}{|C|} \sum_{\substack{(\alpha_0,\ldots,\alpha_b) \\ \alpha_0,\ldots,\alpha_b \geq 0 \\ \alpha_0+\cdots+\alpha_b=n}} A_{(\alpha_0,\ldots,\alpha_b)} \prod_{j=0}^{b} (\vartheta_j^{(b,\ell)}(z))^{\alpha_j}.$$

$\square$

**Example 3.13.** Let $C$ be the byte error-control code over $R_2$ as stated in Example 2.4 generated by

$$\langle (1,0,0,u,v,1+u), (0,u,0,u+v,uv,u), (uv,0,uv,uv,0,uv) \rangle.$$

This code has type $(16)^1(8)^1(2)^1$ and hence $|C| = 256$. The dual code of $C$ is a byte error-control code of length 6 over $R_2$ and it has $16^4 = 65536$ codewords.

The Hamming weight distribution vectors of the codewords of $C$, the number of codewords, and polynomials $\vartheta_j^{(3,16)}(z)$ for $b=3$ and $t=2$ are shown in Tables I and II for the necessary computations to apply Theorem 3.12.

**Table I**
Hamming weight distribution vectors of the codewords in $C$ and the number of codewords.

| $(\alpha_0,\alpha_1,\alpha_2,\alpha_3)$ | number |
|---|---|
| $(2,0,0,0)$ | 1 |
| $(0,2,0,0)$ | 5 |
| $(0,0,2,0)$ | 26 |
| $(0,0,0,2)$ | 64 |
| $(1,1,0,0)$ | 2 |
| $(1,0,1,0)$ | 1 |
| $(1,0,0,1)$ | 1 |
| $(0,1,1,0)$ | 19 |
| $(0,1,0,1)$ | 31 |
| $(0,0,1,1)$ | 106 |

**Table II**
Polynomials $\vartheta_j^{(3,16)}(z)$ for $t=2$, $b=3$ and $\ell=16$.

| |
|---|
| $\vartheta_0^{(3,16)}(z) = 1 + 720z + 3375z^2$ |
| $\vartheta_1^{(3,16)}(z) = 1 + 224z - 225z^2$ |
| $\vartheta_2^{(3,16)}(z) = 1 - 16z + 15z^2$ |
| $\vartheta_3^{(3,16)}(z) = 1 - z^2$ |

According to the expression of $W(z)$ and Table I, we obtain the $m$-spotty Hamming weight enumerator of $C$ as

$$W(z) = 1 + 10z + 183z^2 + 214z^3 + 104z^6.$$

By applying Theorem 3.12 and Table II, we obtain (here for convenience, we write $\vartheta_j^{(b,\ell)}(z) = \vartheta_j(z)$)

$$\begin{aligned} W^\perp(z) &= \frac{1}{|C|} \sum_{\alpha_0+\alpha_1+\alpha_2+\alpha_3=2} A_{(\alpha_0,\alpha_1,\alpha_2,\alpha_3)} \prod_{j=0}^{3} (\vartheta_j(z))^{\alpha_j} \\ &= \frac{1}{256}\big[(\vartheta_0(z))^2 + 5(\vartheta_1(z))^2 + 26(\vartheta_2(z))^2 + 64(\vartheta_3(z))^2 + 2\vartheta_0(z)\vartheta_1(z) + \vartheta_0(z)\vartheta_2(z) \\ &\quad + \vartheta_0(z)\vartheta_3(z) + 19\vartheta_1(z)\vartheta_2(z) + 106\vartheta_2(z)\vartheta_3(z) + 31\vartheta_1(z)\vartheta_3(z)\big] \\ &= 1 + 60z + 4014z^2 + 21932z^3 + 39529z^4. \end{aligned}$$

## 4 MaWillians type identity for joint $m$-Spotty Hamming weight enumerator over finite commutative Frobenius rings

Sharma et al. defined and studied joint $m$-spotty Hamming weight enumerator for a pair of byte error-control codes over the ring of integers modulo $\ell$ ($\ell$ is an integer) and over arbitrary finite fields in [8]. In this section, we entend their results to arbitrary finite commutative Frobenius rings.

**Lemma 4.1**. Let $t$ be a fixed positive integer, and $a$, $b$ be arbitrary nonnegative integers. If $\bar{a}$ and $\bar{b}$, respectively, are the least nonnegative residues of $a$ and $b$ modulo $t$, then we have

$$\left\lceil \frac{a+b}{t} \right\rceil = \begin{cases} \lfloor \frac{a}{t} \rfloor + \lfloor \frac{b}{t} \rfloor, & \text{if } \bar{a} + \bar{b} = 0; \\ \lfloor \frac{a}{t} \rfloor + \lfloor \frac{b}{t} \rfloor + 1, & \text{if } 0 < \bar{a} + \bar{b} \leq t; \\ \lfloor \frac{a}{t} \rfloor + \lfloor \frac{b}{t} \rfloor + 2, & \text{if } t < \bar{a} + \bar{b} \leq 2t - 2, \end{cases}$$

where $\lfloor x \rfloor$ denotes the floor of $x$ for any real number $x$, i.e., the smallest integer number less than or equal to $x$.

*Proof.* The proof is trivial. $\square$

**Definition 4.2**. Define the functions $f_{01}$, $f_{10}$ and $f_{11}$ for each pair of vectors $\mathbf{u}' = (u'_1, u'_2, \cdots, u'_b)$, $\mathbf{v}' = (v'_1, v'_2, \cdots, v'_b)$ in $R_k^b$, as follows:

- $f_{01}$ is the number of $j$'s such that $u'_j = 0$ and $v'_j \neq 0$;
- $f_{10}$ is the number of $j$'s such that $u'_j \neq 0$ and $v'_j = 0$;
- $f_{11}$ is the number of $j$'s such that $u'_j \neq 0$ and $v'_j \neq 0$;

Note that for each $\mathbf{u}', \mathbf{v}' \in R_k^b$, then $f_{10}(\mathbf{u}', \mathbf{v}') + f_{11}(\mathbf{u}', \mathbf{v}') = w_H(\mathbf{u}')$, $f_{01}(\mathbf{u}', \mathbf{v}') + f_{11}(\mathbf{u}', \mathbf{v}') = w_H(\mathbf{v}')$ where $w_H(\mathbf{u}')$ and $w_H(\mathbf{v}')$ are the Hamming weights of $\mathbf{u}$ and $\mathbf{v}$, respectively.

**Definition 4.3**. We define the functions $J$, $K$, $L$: $R_k^b \times R_k^b \to \mathbb{Z}$ as follows:

$$J(\mathbf{u}', \mathbf{v}') = \begin{cases} \lfloor \frac{f_{01}(\mathbf{u}', \mathbf{v}')}{t} \rfloor, & \text{if } \overline{f_{01}(\mathbf{u}', \mathbf{v}')} + \overline{f_{11}(\mathbf{u}', \mathbf{v}')} = 0; \\ \lfloor \frac{f_{01}(\mathbf{u}', \mathbf{v}')}{t} \rfloor + 1, & \text{if } 0 < \overline{f_{01}(\mathbf{u}', \mathbf{v}')} + \overline{f_{11}(\mathbf{u}', \mathbf{v}')} \leq t; \\ \lfloor \frac{f_{01}(\mathbf{u}', \mathbf{v}')}{t} \rfloor + 2, & \text{if } t < \overline{f_{01}(\mathbf{u}', \mathbf{v}')} + \overline{f_{11}(\mathbf{u}', \mathbf{v}')} \leq 2t - 2, \end{cases}$$

$$K(\mathbf{u}', \mathbf{v}') = \begin{cases} \lfloor \frac{f_{10}(\mathbf{u}', \mathbf{v}')}{t} \rfloor, & \text{if } \overline{f_{10}(\mathbf{u}', \mathbf{v}')} + \overline{f_{11}(u', v')} = 0; \\ \lfloor \frac{f_{10}(\mathbf{u}', \mathbf{v}')}{t} \rfloor + 1, & \text{if } 0 < \overline{f_{10}(\mathbf{u}', \mathbf{v}')} + \overline{f_{11}(u', v')} \leq t; \\ \lfloor \frac{f_{10}(\mathbf{u}', \mathbf{v}')}{t} \rfloor + 2, & \text{if } t < \overline{f_{10}(\mathbf{u}', \mathbf{v}')} + \overline{f_{11}(\mathbf{u}', \mathbf{v}')} \leq 2t - 2, \end{cases}$$

$$L(\mathbf{u}', \mathbf{v}') = \left\lfloor \frac{f_{11}(\mathbf{u}', \mathbf{v}')}{t} \right\rfloor \text{ for every } \mathbf{u}', \mathbf{v}' \in R_k^b.$$

Applying Lemma 4.1, it is easy to see that

$$J(\mathbf{u}', \mathbf{v}') + L(\mathbf{u}', \mathbf{v}') = \left\lceil \frac{f_{01}(\mathbf{u}', \mathbf{v}') + f_{11}(\mathbf{u}', \mathbf{v}')}{t} \right\rceil = \left\lceil \frac{w_H(\mathbf{v}')}{t} \right\rceil = w_M(\mathbf{v}') \text{ and}$$

$$K(\mathbf{u}', \mathbf{v}') + L(\mathbf{u}', \mathbf{v}') = \left\lceil \frac{f_{10}(\mathbf{u}', \mathbf{v}') + f_{11}(\mathbf{u}', \mathbf{v}')}{t} \right\rceil = \left\lceil \frac{w_H(\mathbf{u}')}{t} \right\rceil = w_M(\mathbf{u}')$$

From the disscussion above, we can easily get the following proposition.

**Proposition 4.4**. There exist functions: $J, K, L : R_k^{bn} \times R_k^{bn} \longrightarrow \mathbb{Z}$, satisfying the following:

$$J(\mathbf{u}, \mathbf{v}) + L(\mathbf{u}, \mathbf{v}) = w_M(\mathbf{v}), K(\mathbf{u}, \mathbf{v}) + L(\mathbf{u}, \mathbf{v}) = w_M(\mathbf{u})$$

for all $\mathbf{u} = (\mathbf{u}_1, \mathbf{u}_2, \cdots, \mathbf{u}_n), \mathbf{v} = (\mathbf{v}_1, \mathbf{v}_2, \cdots, \mathbf{v}_n) \in R_k^{bn}$ with $\mathbf{u}_i$'s and $\mathbf{v}_i$'s in $R_k^b$.

Now we are ready to define joint $m$-spotty Hamming weight enumerator for a pair of byte error-control codes over $R_k$.

**Definition 4.5.** Let $C$ and $D$ be byte error-control codes of length $bn$ and byte length $b$ over $R_k$. Then the joint $m$-spotty Hamming weight enumerator of the codes $C$ and $D$ is defined as

$$\mathcal{J}_{(C,D)}(x,y,z) = \sum_{\mathbf{u} \in C} \sum_{\mathbf{v} \in D} x^{J(\mathbf{u},\mathbf{v})} y^{K(\mathbf{u},\mathbf{v})} z^{L(\mathbf{u},\mathbf{v})}.$$

The following theorem shows that the joint $m$-spotty weight enumerator generalizes $m$-spotty weight neumerator just like the joint probability density function generalizes single probability density function.

**Theorem 4.6.** The joint $m$-spotty Hamming weight enumerator $\mathcal{J}_{C,D}(x,y,z)$ of byte error-control codes $C$ and $D$ over $R_k$ satisfies the following properties:

(i) $\mathcal{J}_{(C,D)}(1,1,1) = |C||D|$,
(ii) $\mathcal{J}_{(D,C)}(x,y,z) = \mathcal{J}_{(C,D)}(y,x,z)$,
(iii) $W_C(z) = \frac{1}{|D|} \mathcal{J}_{(C,D)}(1,z,z)$, where $W_C(z)$ is the $m$-spotty Hamming weight enumerator of $C$,
(iv) $W_D(z) = \frac{1}{|C|} \mathcal{J}_{(C,D)}(z,1,z)$, where $W_D(z)$ is the $m$-spotty Hamming weight enumerator of $D$.

*Proof.* The proof is similar to that of Theorem 12 in [8]. □

For our purpose, we need the following definitions.

**Definition 4.7.** Let $t, \nu, \mu$ be the integers satisfying $1 \leq t \leq b$ and $0 \leq \nu, \mu \leq b$, and let $\delta$ be an integer satisfying $\nu + \mu - b \leq \delta \leq \min\{\nu, \mu\}$. For an integer $p$ $(0 \leq p \leq b)$, let $\mathcal{A}_p$ be the set of all 4-tuple $\alpha = (\alpha_1, \alpha_2, \alpha_3, \alpha_4)$ of nonnegative integers $\alpha_i$'s satisfying $\alpha_1 + \alpha_2 + \alpha_3 + \alpha_4 = p$ with $0 \leq \alpha_1 \leq \delta$, $0 \leq \alpha_2 \leq \mu - \delta$, $0 \leq \alpha_3 \leq \nu - \delta$, $0 \leq \alpha_4 \leq b + \delta - \nu - \delta$. Then we define the polynomial $G_{\nu,\mu}^{\delta}(x,y,z)$ as

$$\sum_{p=0}^{b} \sum_{p} g_p(\alpha) x^{\lfloor \frac{\mu - \alpha_1 - \alpha_2}{t} \rfloor + \theta_p^{(\alpha)}} y^{\lfloor \frac{p - \alpha_1 - \alpha_2}{t} \rfloor + \eta_p^{(\alpha)}} z^{\lfloor \frac{\alpha_1 + \alpha_2}{t} \rfloor}$$

where for each $p$ $(0 \leq p \leq b)$, the summation $\sum_p$ runs over the set $\mathcal{A}_p$; and further for each $\alpha \in \mathcal{A}_p$, the coefficient $g_p(\alpha)$ is given by

$$\binom{\delta}{\alpha_1}\binom{\mu - \delta}{\alpha_2}\binom{\nu - \delta}{\alpha_3}\binom{b + \delta - \mu - \nu}{\alpha_4}(-1)^{\alpha_1 + \alpha_3}(\ell - 1)^{p - \alpha_1 - \alpha_2}.$$

and the number $\theta_p^{(\alpha)}, \eta_p^{(\alpha)}$ are given by

$$\theta_p^{(\alpha)} = \begin{cases} 0, & \text{if } \overline{\mu - \alpha_1 - \alpha_2} + \overline{\alpha_1 + \alpha_2}; \\ 1, & \text{if } 0 < \overline{\mu - \alpha_1 - \alpha_2} + \overline{\alpha_1 + \alpha_2} \leq t; \\ 2, & \text{if } t < \overline{\mu - \alpha_1 - \alpha_2} + \overline{\alpha_1 + \alpha_2} \leq 2t - 2, \end{cases}$$

$$\eta_p^{(\alpha)} = \begin{cases} 0, & \text{if } \overline{p - \alpha_1 - \alpha_2} + \overline{\alpha_1 + \alpha_2}; \\ 1, & \text{if } 0 < \overline{p - \alpha_1 - \alpha_2} + \overline{\alpha_1 + \alpha_2} \leq t; \\ 2, & \text{if } t < \overline{p - \alpha_1 - \alpha_2} + \overline{\alpha_1 + \alpha_2} \leq 2t - 2, \end{cases}$$

**Definition 4.8.** For each $a \in \mathbb{Z}_2^4$, let $[a]_i (1 \leq i \leq b)$ denote the $i$-th component of $a$. Then for any 4-tuple $(\ell_1, \ell_2, \ell_3, \ell_4)$ over the set $\{0, 1, *\}$, we define $T_{\ell_1 \ell_2 \ell_3 \ell_4}$ as the set of all 4-tuple $a \in \mathbb{Z}_2^4$ satisfying $[a]_i = \ell_i$ if $\ell_i$ is either 0 or 1; and $[a]_i$ runs over $\mathbb{Z}_2$ if $\ell_i = *$.

**Definition 4.9.** Let $t(1 \leq t \leq b)$ be an integer and let $\mu, \nu, \delta$ be the integers satisfying $0 \leq \mu, \nu, \delta \leq b$. For integers $p, q(0 \leq p, q \leq b)$, let $\mathcal{B}_{pq}$ be the set of all tuples $\alpha = (\alpha_a : a \in \mathbb{Z}_2^4)$ of nonnegative integers $\alpha_a$'s satisfying the following:

$$\sum_{a \in T_{1**1}} \alpha_a = \delta, \sum_{a \in T_{1***}} \alpha_a = \nu, \sum_{a \in T_{***1}} \alpha_a = \mu, \sum_{a \in \mathbb{Z}_2^4} \alpha_a = b, \sum_{a \in T_{*1**}} \alpha_a = p, \sum_{a \in T_{**1*}} \alpha_a = q.$$

Then we define the polynomial $H_{\mu,\nu}^{(\delta)}(x,y,z)$ as

$$\sum_{p,q=0}^{b} \sum_{p,q} h_{pq}(\alpha) x^{\left\lfloor \frac{q-\psi_{pq}^{(\alpha)}}{t} \right\rfloor + \zeta_{pq}^{(\alpha)}} y^{\left\lfloor \frac{p-\psi_{pq}^{(\alpha)}}{t} \right\rfloor + \omega_{pq}^{(\alpha)}} z^{\left\lfloor \frac{\psi_{pq}^{(\alpha)}}{t} \right\rfloor},$$

where for each $p, q (0 \leq p, q \leq b)$, the summation $\sum_{p,q}$ runs over the set $\mathcal{B}_{pq}$; and further for each tuple $\alpha \in \mathcal{B}_{pq}$, the coefficient $h_{pq}(\alpha)$ is given by

$$\frac{\delta!(\nu-\delta)!(\mu-\delta)!(b+\delta-\nu-\mu)!}{\prod_{a \in \mathbb{Z}_2^4} \alpha_a!} (-1)^{\phi_{pq}^{(\alpha)}} (\ell-1)^{p+q-\phi_{pq}(\alpha)}$$

with $\phi_{pq}^{(\alpha)} = \sum_{a \in T_{11**}} \alpha_a + \sum_{a \in T_{**11}} \alpha_a, \psi_{pq}^{(\alpha)} = \sum_{a \in T_{*11*}} \alpha_a$ and the numbers

$$\zeta_{pq}^{(\alpha)} = \begin{cases} 0, & \text{if } \overline{q - \psi_{pq}^{(\alpha)}} + \overline{\psi_{pq}^{(\alpha)}}; \\ 1, & \text{if } 0 < \overline{q - \psi_{pq}^{(\alpha)}} + \overline{\psi_{pq}^{(\alpha)}} \leq t; \\ 2, & \text{if } t < \overline{q - \psi_{pq}^{(\alpha)}} + \overline{\psi_{pq}^{(\alpha)}} \leq 2t-2, \end{cases} \quad (3)$$

$$\omega_{pq}^{(\alpha)} = \begin{cases} 0, & \text{if } \overline{p - \psi_{pq}^{(\alpha)}} + \overline{\psi_{pq}^{(\alpha)}}; \\ 1, & \text{if } 0 < \overline{p - \psi_{pq}^{(\alpha)}} + \overline{\psi_{pq}^{(\alpha)}} \leq t; \\ 2, & \text{if } t < \overline{p - \psi_{pq}^{(\alpha)}} + \overline{\psi_{pq}^{(\alpha)}} \leq 2t-2, \end{cases} \quad (4)$$

If, for $1 \leq i \leq n$, $j_i$ is the Hamming weight of $i$-th byte $\mathbf{u}_i = (u_{i1}, u_{i2}, \cdots, u_{ib})$ of the vector $u = (\mathbf{u}_1, \mathbf{u}_2, \cdots, \mathbf{u}_n) \in R_k^{bn}$, then the vector $(j_1, j_2, \cdots, j_n)$ is called the Hamming weight distribution vector of $\mathbf{u}$ and is denoted by $w_D(\mathbf{u})$. Let $j = (j_1, j_2, \cdots, j_n), k = (k_1, k_2, \cdots, k_n)$ and $\delta = (\delta_1, \delta_2, \cdots, \delta_n)$ be the Hamming weight distribution vectors of $\mathbf{u}, \mathbf{v}$ and $\mathbf{w}$, where $\mathbf{u}, \mathbf{v}, \mathbf{w} \in R_k^{bn}$ and $0 \leq j_i, k_i, \delta_i \leq b$ for each $i$, we define the polynomials

$$G_{j,k}^{(\delta)}(x,y,z) = \prod_{i=1}^{n} G_{j_i,k_i}^{(\delta_i)}(x,y,z), H_{j,k}^{(\delta)}(x,y,z) = \prod_{i=1}^{n} H_{j_i,k_i}^{(\delta_i)}(x,y,z).$$

We also define $j \vee k = (j_1 \vee k_1, j_2 \vee k_2, \cdots, j_n \vee k_n)$, where $j_i \vee k_i = \min\{j_i, k_i\}$ for each $i$. Further, let $\mathbf{v} = (\mathbf{v}_1, \mathbf{v}_2, \cdots, \mathbf{v}_n)$ where $\mathbf{v}_i = (v_{i1}, v_{i2}, \cdots, v_{ib}) \in R_k^b$. Then we define a vector $\mathbf{u} \cap \mathbf{v} \in \mathbb{Z}_2^{bn}$ as $\mathbf{u} \cap \mathbf{v} = (\mathbf{u}_1 \cap \mathbf{v}_1, \mathbf{u}_2 \cap \mathbf{v}_2, \cdots, \mathbf{u}_n \cap \mathbf{v}_n)$, where for each $i$, $\mathbf{u}_i \cap \mathbf{v}_i = (u_{i1} \cap v_{i1}, u_{i2} \cap v_{i2}, \cdots, u_{ib} \cap v_{ib})$ with each $u_{ij} \cap v_{ij}$ given by

$$u_{ij} \cap v_{ij} = \begin{cases} 1 & \text{if } u_{ij} \neq 0 \text{ and } v_{ij} \neq 0; \\ 0 & \text{othwise} \end{cases}$$

it is easy to see that $0 \leq w_H(\mathbf{u}_i \cap \mathbf{v}_i) \leq j_i \vee k_i$ for each $i$, so that $w_D(\mathbf{u} \cap \mathbf{v})$ varies from 0 to $j \vee k$.

**Theorem 4.10**. Let $C$ and $D$ be byte error-control codes of length $bn$ and byte length $b$ over $R_k$. If $A_\delta(j;k)$ is the number of pairs $(\mathbf{u}, \mathbf{v})$ of codewords $\mathbf{u} \in C$ and $\mathbf{v} \in D$ having $w_D(\mathbf{u}) = j, w_D(\mathbf{v}) = k$ and $w_D(\mathbf{u} \cap \mathbf{v}) = \delta$, then we have

(i) $\mathcal{J}_{C^\perp, D}(x,y,z) = \frac{1}{|C|} \sum_{j,k} \sum_{\delta=0}^{j \vee k} A_\delta(j;k) G_{j,k}^{(\delta)}(x,y,z),$

(ii) $\mathcal{J}_{C, D^\perp}(x,y,z) = \frac{1}{|D|} \sum_{j,k} \sum_{\delta=0}^{j \vee k} A_\delta(j;k) G_{k,j}^{(\delta)}(y,x,z),$

(iii) $\mathcal{J}_{C^\perp, D^\perp}(x,y,z) = \frac{1}{|C||D|} \sum_{j,k} \sum_{\delta=0}^{j \vee k} A_\delta(j;k) H_{j,k}^{(\delta)}(y,x,z),$

where the summation $\sum_{j,k}$ runs over all $n$-tuples $j = (j_1, j_2, \cdots, j_n)$ and $k = (k_1, k_2, \cdots, k_n)$ satisfying $0 \leq j_i, k_i \leq b$, and the polynomials $G_{j,k}^{(\delta)}(x,y,z)$'s and $H_{j,k}^{(\delta)}(x,y,z)$'s are defined in Definition 4.9.

*Proof.* The proof is similar to those of Theorem 3.12 and Theorem 20 in [8]. □

**Example 4.11**. Let $C$ be the byte error-control code over $R_2$ generated by the set

$$\{(1,0,0,u,v,1,0,0,u), (0,0,uv,uv,0,0,0,uv,uv)\}.$$

Its length is 9 and byte length is 3. It is easy to check that the generators are independent, hence the code has type $(16)^1(2)^1$ and $|C| = 32$. Its dual code $C^\perp$, which is also a byte error-control code over $R_2$ of length 9, contains $16^9 \div 32 = 2147483648$ codewords. Let $D$ be a linear code of length 9 and byte length 3, generated by $(0,0,uv,uv,0,0,0,uv,uv)$. Since $|C^\perp| = 2147483648$ is large and $|D| = 2$ is small, we apply Theorem 4.10 (i) to obtain joint $m$-spotty weight enumerator of the codes $C^\perp$ and $D$. For this, we need to compute Hamming weight distribution vectors of the codewords in $C$, which are given in Table IV. It is easy to see that the codewords $\mathbf{u} = (1,0,0,u,v,1,0,0,u) \in C$ and the unique nonzero $\mathbf{v} = (uv,0,uv,0,uv,0,0,0,uv) \in D$ having Hamming weight distribution vectors as $j = (1,3,1)$ and $k = (2,1,1)$, respectively, have $\delta = (1,1,1)$ and they contribute

$$G_{j,k}^{(\delta)}(x,y,z) = G_{1,2}^{(1)}(x,y,z) G_{3,1}^{(1)}(x,y,z) G_{1,1}^{(1)}(x,y,z)$$

to the joint $m$-spotty Hamming weight enumerator, where by Definition 4.7,

$G_{1,2}^{(1)}(x,y,z) = x + 239xy - 225yz - 15z$, $G_{3,1}^{(1)}(x,y,z) = x - xy^2$ and $G_{1,1}^{(1)}(x,y,z) = x + 224xy - 225xy^2$.

Working in a similar way, we obtain the contributing polynomials for each triplet of $j, k$ and $\delta$, which are given in Table III (here for convenience, we write $G_{j,k}^{(\delta)}(x,y,z) = G_{j,k}^{(\delta)}$). Now by Theorem 4.10 (i), the joint $m$-spotty Hamming weight enumerator of $C$ and $D$ is given by

$$\begin{aligned}
\mathcal{J}_{C^\perp, D}(x,y,z) &= \frac{1}{|C|} \sum_{j,k} \sum_{\delta=0}^{j \vee k} A_\delta(j;k) G_{j,k}^{(\delta)}(x,y,z) \\
&= 171174300 x^3 y^5 + 79546455 x^3 y^4 + 9241586 x^3 y^3 + 84136 x^3 y^2 + 370 x^3 y + x^3 \\
&\quad + 1206082125 x^2 y^5 z + 589195350 x^2 y^4 z + 88403600 x^2 y^3 z + 3748010 x^2 y^2 z \\
&\quad + 7715 x^2 yz + 1206082125 y^6 + 760369650 y^5 + 167950055 y^4 + 12989596 y^3 \\
&\quad + 91851 y^2 + 370 y + 1.
\end{aligned}$$

## 5 MaWillians type identity for split $m$-Spotty Hamming weight enumerator over Finite commutative Frobenius rings

In this section, we prove that the results in [8] still hold over arbitrary finite commutative Frobenius rings. We first recall split $m$-spotty Hamming weight enumerator for a byte error-control code over $R_k$ as follows:

**Definition 5.1**. Let $C$ be a byte error-control code of length $bn$ over $R_k$ with byte length $b$. Then the split $m$-spotty Hamming weight enumerator of the code $C$, denoted by $S_C(x_i, y_i : i = 1, 2, \cdots, n)$, is defined as

$$\sum_{(\mathbf{u_1}, \mathbf{u_2}, \cdots, \mathbf{u_n}) \in C} \left( \prod_{i=1}^n x_i^{\lceil b/t \rceil - w_M(\mathbf{u}_i)} y_i^{w_M(\mathbf{u}_i)} \right).$$

We recall that if $j_i (1 \leq i \leq n)$ is the Hamming weight of the $i$-th byte $\mathbf{u}_i$ of $\mathbf{u} = (\mathbf{u_1}, \mathbf{u_2}, \cdots, \mathbf{u_n})$, then the vector $(j_1, j_2, \cdots, j_n)$ is called the Hamming weigth distribution vector of $\mathbf{u}$. If $A(j_1, j_2, \cdots, j_n)$ denotes the number of codewords in $C$ having Hamming weight distribution vector as

| $j$ | $k$ | $\delta$ | $A_\delta(j,k)$ | $G_{j,k}^{(\delta)}(x,y,z)$ |
|---|---|---|---|---|
| (0,0,0) | (0,0,0) | (0,0,0) | 1 | $G_{0,0}^{(0)}G_{0,0}^{(0)}G_{0,0}^{(0)}$ |
| (0,0,0) | (2,1,1) | (0,0,0) | 1 | $G_{0,2}^{(0)}G_{0,1}^{(0)}G_{0,1}^{(0)}$ |
| (1,1,0) | (0,0,0) | (0,0,0) | 1 | $G_{1,0}^{(0)}G_{1,0}^{(0)}G_{0,0}^{(0)}$ |
| (1,1,0) | (2,1,1) | (1,0,0) | 1 | $G_{1,2}^{(1)}G_{1,1}^{(0)}G_{0,1}^{(0)}$ |
| (1,1,2) | (0,0,0) | (0,0,0) | 1 | $G_{1,0}^{(0)}G_{1,0}^{(0)}G_{2,0}^{(0)}$ |
| (1,1,2) | (2,1,1) | (1,0,1) | 1 | $G_{1,2}^{(1)}G_{1,1}^{(0)}G_{2,1}^{(1)}$ |
| (1,2,0) | (0,0,0) | (0,0,0) | 2 | $G_{1,0}^{(0)}G_{2,0}^{(0)}G_{0,0}^{(0)}$ |
| (1,2,0) | (2,1,1) | (1,1,0) | 2 | $G_{1,2}^{(1)}G_{2,1}^{(1)}G_{0,1}^{(0)}$ |
| (1,2,1) | (0,0,0) | (0,0,0) | 2 | $G_{1,0}^{(0)}G_{2,0}^{(0)}G_{1,0}^{(0)}$ |
| (1,2,1) | (2,1,1) | (1,0,1) | 2 | $G_{1,2}^{(1)}G_{2,1}^{(0)}G_{1,1}^{(1)}$ |
| (1,3,1) | (0,0,0) | (0,0,0) | 10 | $G_{1,0}^{(0)}G_{3,0}^{(0)}G_{1,0}^{(0)}$ |
| (1,3,1) | (2,1,1) | (1,1,1) | 10 | $G_{1,2}^{(1)}G_{3,1}^{(1)}G_{1,1}^{(1)}$ |
| (2,1,1) | (0,0,0) | (0,0,0) | 2 | $G_{2,0}^{(0)}G_{1,0}^{(0)}G_{1,0}^{(0)}$ |
| (2,1,1) | (2,1,1) | (2,0,0) | 2 | $G_{2,2}^{(2)}G_{1,1}^{(0)}G_{1,1}^{(0)}$ |
| (2,2,1) | (0,0,0) | (0,0,0) | 2 | $G_{2,0}^{(0)}G_{2,0}^{(0)}G_{1,0}^{(0)}$ |
| (2,2,1) | (2,1,1) | (2,1,0) | 2 | $G_{2,2}^{(2)}G_{2,1}^{(1)}G_{1,1}^{(0)}$ |
| (2,2,2) | (0,0,0) | (0,0,0) | 1 | $G_{2,0}^{(0)}G_{2,0}^{(0)}G_{2,0}^{(0)}$ |
| (2,2,2) | (2,1,1) | (2,0,1) | 1 | $G_{2,2}^{(2)}G_{2,1}^{(0)}G_{2,1}^{(1)}$ |
| (2,3,2) | (0,0,0) | (0,0,0) | 10 | $G_{2,0}^{(0)}G_{3,0}^{(0)}G_{2,0}^{(0)}$ |
| (2,3,2) | (2,1,1) | (2,1,1) | 10 | $G_{2,2}^{(2)}G_{3,1}^{(1)}G_{2,1}^{(1)}$ |

Table III Contributing polynomials to the $m$-spotty Hamming weight enumerator

$G_{0,0}^{(0)} = 1 + 720y + 3375y^2,$
$G_{0,1}^{(0)} = x + 720xy + 3375xy^2,$
$G_{0,2}^{(0)} = x + 495xy + 225z + 3375yz,$
$G_{1,0}^{(0)} = 1 + 224y - 225y^2,$
$G_{1,1}^{(0)} = G_{1,1}^{(1)} = x + 224xy - 225xy^2,$
$G_{1,2}^{(1)} = x + 239xy - 225yz - 15z,$
$G_{2,0}^{(0)} = 1 - 16y + 15y^2,$
$G_{2,1}^{(0)} = G_{2,1}^{(1)} = x - 16xy + 15xy^2,$
$G_{2,2}^{(2)} = x + z + 15yz - 17xy,$
$G_{3,0}^{(0)} = 1 - y^2,$
$G_{3,1}^{(1)} = x - xy^2.$

$(j_1, j_2, \cdots, j_n)$, then the split $m$-spotty Hamming weight enumerator $S_C(x_i, y_i : i = 1, 2, \cdots, n)$ of the code $C$ can be rewritten as

$$\sum_{(\mathbf{u}_1, \mathbf{u}_2, \cdots, \mathbf{u}_n) \in C} A(j_1, j_2, \cdots, j_n) \prod_{i=1}^{n} x_i^{\lceil b/t \rceil - \lceil j_i/t \rceil} y_i^{\lceil j_i/t \rceil},$$

where the summation runs over all $n$-tuples $(j_1, j_2, \cdots, j_n)$ satisfying $0 \leq j_i \leq b$ for each $i$, $1 \leq i \leq n$.

**Theorem 5.2.** Let $C$ be a byte error-control code of length $bn$ over $R_k$ with byte length $b$ and split $m$-spotty Hamming weight enumerator $S_C(x_i, y_i : i = 1, 2, \cdots, n)$. Then the split $m$-spotty Hamming weight enumerator of the dual code $C^\perp$ over $R_k$ is given by

$$S_{C^\perp}(x_i, y_i : i = 1, 2, \cdots, n) = \frac{1}{|C|} \sum_{(j_1, j_2, \cdots, j_n)} A(j_1, j_2, \cdots, j_n) \prod_{i=1}^{n} g_{j_i}^{(t)}(x_i, y_i),$$

where the summation runs over all $n$-tuples $(j_1, j_2, \cdots, j_n)$ satisfying $0 \leq j_i \leq b$ for $1 \leq i \leq n$, and the polynomials $g_{j_i}^{(t)}(x_i, y_i)$ are definied by

$$g_{j_i}^{(t)}(x_i, y_i) = \sum_{p=0}^{b} K_p(j_i) x^{\lceil b/t \rceil - \lceil p/t \rceil} y^{\lceil p/t \rceil}, \tag{5}$$

where for each $p$, the polynomial $K_p(j_i) = \sum_{a=0}^{p} (-1)^a (\ell-1)^{p-a} \binom{j_i}{a} \binom{b-j_i}{p-a}$ is the well-known Krawtchouk polynomial. (Here, we assume that $\binom{e}{f} = 0$ when $f < 0$ or $f > e$.)

*Proof.* The proof is similar to those of Theorem 3.12 and Theorem 25 in [8]. □

**Theorem 5.3.** Let $C, D$ be byte error-control codes of length $bn$ over $R_k$ with byte length $b$ and $m$-spotty Hamming weight enumerator $W_C(z), W_D(z)$ and split $m$-spotty Hamming weight enumerators $S_C(x_i, y_i : i = 1, 2, \cdots, n), S_D(x_i, y_i : i = 1, 2, \cdots, n)$, respectively. Then

(i) the direct sum $C \oplus D = (\mathbf{u}|\mathbf{v} : \mathbf{u} \in C, \mathbf{v} \in D)$ has $m$-spotty Hamming weight enumerator $W_C(z)W_D(z)$ and split $m$-spotty Hamming weight enumerator

$$S_C(x_i, y_i : i = 1, 2, \cdots, n) S_D(X_i, Y_i : i = 1, 2, \cdots, n).$$

(ii) assuming $n$ even, the code $C || D = \{(\mathbf{u}'|\mathbf{v}'|\mathbf{u}''|\mathbf{v}'') : \mathbf{u} = (\mathbf{u}'|\mathbf{u}'') \in C, \mathbf{v} = (\mathbf{v}'|\mathbf{v}'') \in D\}$ (where $\mathbf{u}$ and $\mathbf{v}$ have each been broken into two equal halves) has $m$-spotty Hamming weight enumerator $W_C(z)W_D(z)$ and split $m$-spotty Hamming weight enumerator

$$S_C(x_i, y_i; X_i, Y_i : i = 1, 2, \cdots, n/2) S_D(x_i, y_i; X_i, Y_i : i = (n/2) + 1, \cdots, n).$$

*Proof.* The proof is similar to that of Theorem 28 in [8]. □

**Example 5.4.** Let $C$ be the byte error-control code as defined in Example 4.11. Here we apply Theorem 5.2 to compute split $m$-spotty Hamming weight enumerator of the code $C^\perp$. For this, first we need to compute the Hamming weight distribution vectors for the codewords of $C$.

It is easy to see that the codeword $(1, 0, 0, u, v, 1, 0, 0, u) \in C$ has Hamming weight distribution vector as $(1, 3, 1)$ and it contributes the polynomial $g_1^{(2)}(x_1, y_1) g_3^{(2)}(x_2, y_2) g_1^{(2)}(x_3, y_3)$ to the split $m$-spotty Hamming weight enumerator, where by (5), we have $g_1^{(2)}(x_i, y_i) = x_i^2 + 224 x_i y_i - 225 y_i^2, g_3^{(2)}(x_i, y_i) = x_i^2 - y_i^2$ for each $i$.

Working similarly, we obtain the Hamming weight distribution vector and the contributing polynomials for other codewords in $C$, which are given in Table IV. Therefore by Theorem 5.2, the split $m$-spotty Hamming weight enumerator of the code $C$ is given by

$$\begin{aligned}
S_C^\perp(x_i, y_i : i = 1, 2, 3) &= \frac{1}{|C|} \sum_{(j_1, j_2, j_3)} \prod_{i=1}^{3} g_{j_i}^{(2)}(x_i, y_i) \\
&= 2x_1^2 x_2^2 x_3^2 + 392 x_1^2 x_2^2 x_3 y_3 + 630 x_1^2 x_2^2 y_3^2 + 94 x_1^2 x_2 y_2 x_3^2 + 46208 x_1^2 x_2 y_2 x_3 y_3 \\
&\quad + 187170 x_1^2 x_2 y_2 y_3^2 + 160 x_1^2 y_2^2 x_3^2 + 137720 x_1^2 y_2^2 x_3 y_3 + 676200 x_1^2 y_2^2 y_3^2 \\
&\quad + 254 x_1 y_1 x_2^2 x_3^2 + 99328 x_1 y_1 x_2^2 x_3 y_3 + 256770 x_1 y_1 x_2^2 y_3^2 + 37376 x_1 y_1 x_2 y_2 x_3^2 \\
&\quad + 25019392 x_1 y_1 x_2 y_2 x_3 y_3 + 118663680 x_1 y_1 x_2 y_2 y_3^2 + 146690 x_1 y_1 y_2^2 x_3^2 \\
&\quad + 107591680 x_1 y_1 y_2^2 x_3 y_3 + 503159550 x_1 y_1 y_2^2 y_3^2 + 84600 y_1^2 x_2^2 x_3 y_3 \\
&\quad + 606600 y_1^2 x_2^2 y_3^2 + 146850 y_1^2 x_2 y_2 x_3^2 + 107644800 y_1^2 x_2 y_2 x_3 y_3 + 503229150 \\
&\quad y_1^2 x_2 y_2 y_3^2 + 717150 y_1^2 y_2^2 x_3^2 + 514350600 y_1^2 y_2^2 x_3 y_3 + 2412164250 y_1^2 y_2^2 y_3^2.
\end{aligned}$$

## 6 MacWilliams type identity for $m$-spotty Lee weight enumerator over an infinite family of rings

In this section, we concentrate our study on $m$-spotty Lee weight enumerator for byte error-control codes over $R_k = R_{k-1}[u_k]/\langle u_k^2 = 0, u_k u_j = u_j u_k \rangle$. We prove that the results in [9] are still valid over $R_k$. For this, we first recall the definition of Lee weights in $R_k$ as follows:

**Definition 6.1.** For any subset $A \subseteq \{1, 2, \cdots, k\}$, the Lee weight of $u_A$ is defined by

$$w_L(u_A) = 2^{|A|}.$$

In $R_k$, there are precisely $\binom{2^k}{i}$ elements of Lee weight $i$, for $i = 0, 1, \cdots, 2^k$. For example, consider $R_2 = \mathbb{F}_2 + u\mathbb{F}_2 + v\mathbb{F}_2 + uv\mathbb{F}_2$, the Lee weight of elements $1, 1 + u, 1 + v$ and $1 + u + v + uv$ is 1; of elements $u, v, u + v, u + uv, v + uv$ and $u + v + uv$ is 2; of elements $1 + uv, 1 + u + uv, 1 + v + uv$ and $1 + u + v$ is 3; of element $uv$ is 4; of element 0 is 0.

Table IV Contributing polynomials of the codewords

| codewords of $C$ | $(j_1, j_2, j_3)$ | $g_{j_1}^{(2)}(x_1,y_1)g_{j_2}^{(2)}(x_2,y_2)g_{j_3}^{(2)}(x_3,y_3)$ |
|---|---|---|
| $(0,0,0,0,0,0,0,0,0)$ | $(0,0,0)$ | $g_0^{(2)}(x_1,y_1)g_0^{(2)}(x_2,y_2)g_0^{(2)}(x_3,y_3)$ |
| $(8,0,0,0,0,8,0,0,0)$ | $(1,1,0)$ | $g_1^{(2)}(x_1,y_1)g_1^{(2)}(x_2,y_2)g_0^{(2)}(x_3,y_3)$ |
| $(0,0,8,8,0,0,0,8,8)$ | $(1,1,2)$ | $g_1^{(2)}(x_1,y_1)g_1^{(2)}(x_2,y_2)g_2^{(2)}(x_3,y_3)$ |
| $(2,0,0,0,8,2,0,0,0)$ | $(1,2,0)$ | $g_1^{(2)}(x_1,y_1)g_2^{(2)}(x_2,y_2)g_0^{(2)}(x_3,y_3)$ |
| $(10,0,0,0,8,10,0,0,0)$ | $(1,2,0)$ | $g_1^{(2)}(x_1,y_1)g_2^{(2)}(x_2,y_2)g_0^{(2)}(x_3,y_3)$ |
| $(4,0,0,8,0,4,0,0,8)$ | $(1,2,1)$ | $g_1^{(2)}(x_1,y_1)g_2^{(2)}(x_2,y_2)g_1^{(2)}(x_3,y_3)$ |
| $(12,0,0,8,0,12,0,0,8)$ | $(1,2,1)$ | $g_1^{(2)}(x_1,y_1)g_2^{(2)}(x_2,y_2)g_1^{(2)}(x_3,y_3)$ |
| $(1,0,0,2,4,1,0,0,2)$ | $(1,3,1)$ | $g_1^{(2)}(x_1,y_1)g_3^{(2)}(x_2,y_2)g_1^{(2)}(x_3,y_3)$ |
| $(3,0,0,2,12,3,0,0,2)$ | $(1,3,1)$ | $g_1^{(2)}(x_1,y_1)g_3^{(2)}(x_2,y_2)g_1^{(2)}(x_3,y_3)$ |
| $(5,0,0,10,4,5,0,0,10)$ | $(1,3,1)$ | $g_1^{(2)}(x_1,y_1)g_3^{(2)}(x_2,y_2)g_1^{(2)}(x_3,y_3)$ |
| $(6,0,0,8,8,6,0,0,8)$ | $(1,3,1)$ | $g_1^{(2)}(x_1,y_1)g_3^{(2)}(x_2,y_2)g_1^{(2)}(x_3,y_3)$ |
| $(7,0,0,10,12,7,0,0,10)$ | $(1,3,1)$ | $g_1^{(2)}(x_1,y_1)g_3^{(2)}(x_2,y_2)g_1^{(2)}(x_3,y_3)$ |
| $(9,0,0,2,4,9,0,0,2)$ | $(1,3,1)$ | $g_1^{(2)}(x_1,y_1)g_3^{(2)}(x_2,y_2)g_1^{(2)}(x_3,y_3)$ |
| $(11,0,0,2,12,11,0,0,2)$ | $(1,3,1)$ | $g_1^{(2)}(x_1,y_1)g_3^{(2)}(x_2,y_2)g_1^{(2)}(x_3,y_3)$ |
| $(13,0,0,10,4,13,0,0,10)$ | $(1,3,1)$ | $g_1^{(2)}(x_1,y_1)g_3^{(2)}(x_2,y_2)g_1^{(2)}(x_3,y_3)$ |
| $(14,0,0,8,8,14,0,0,8)$ | $(1,3,1)$ | $g_1^{(2)}(x_1,y_1)g_3^{(2)}(x_2,y_2)g_1^{(2)}(x_3,y_3)$ |
| $(15,0,0,10,12,15,0,0,10)$ | $(1,3,1)$ | $g_1^{(2)}(x_1,y_1)g_3^{(2)}(x_2,y_2)g_1^{(2)}(x_3,y_3)$ |
| $(4,0,8,0,0,4,0,8,0)$ | $(2,1,1)$ | $g_2^{(2)}(x_1,y_1)g_1^{(2)}(x_2,y_2)g_1^{(2)}(x_3,y_3)$ |
| $(12,0,8,0,0,12,0,8,0)$ | $(2,1,1)$ | $g_1^{(2)}(x_1,y_1)g_1^{(2)}(x_2,y_2)g_1^{(2)}(x_3,y_3)$ |
| $(6,0,8,0,8,6,0,8,0)$ | $(2,2,1)$ | $g_2^{(2)}(x_1,y_1)g_2^{(2)}(x_2,y_2)g_1^{(2)}(x_3,y_3)$ |
| $(14,0,8,0,8,14,0,8,0)$ | $(2,2,1)$ | $g_3^{(2)}(x_1,y_1)g_2^{(2)}(x_2,y_2)g_1^{(2)}(x_3,y_3)$ |
| $(8,0,8,8,0,8,0,8,8)$ | $(2,2,2)$ | $g_2^{(2)}(x_1,y_1)g_2^{(2)}(x_2,y_2)g_2^{(2)}(x_3,y_3)$ |
| $(1,0,8,10,4,1,0,8,10)$ | $(2,3,2)$ | $g_2^{(2)}(x_1,y_1)g_3^{(2)}(x_2,y_2)g_2^{(2)}(x_3,y_3)$ |
| $(2,0,8,8,8,2,0,8,8)$ | $(2,3,2)$ | $g_2^{(2)}(x_1,y_1)g_3^{(2)}(x_2,y_2)g_2^{(2)}(x_3,y_3)$ |
| $(3,0,8,10,12,3,0,8,10)$ | $(2,3,2)$ | $g_2^{(2)}(x_1,y_1)g_3^{(2)}(x_2,y_2)g_2^{(2)}(x_3,y_3)$ |
| $(5,0,8,2,4,5,0,8,2)$ | $(2,3,2)$ | $g_2^{(2)}(x_1,y_1)g_3^{(2)}(x_2,y_2)g_2^{(2)}(x_3,y_3)$ |
| $(7,0,8,2,12,7,0,8,2)$ | $(2,3,2)$ | $g_2^{(2)}(x_1,y_1)g_3^{(2)}(x_2,y_2)g_2^{(2)}(x_3,y_3)$ |
| $(9,0,8,10,4,9,0,8,10)$ | $(2,3,2)$ | $g_2^{(2)}(x_1,y_1)g_3^{(2)}(x_2,y_2)g_2^{(2)}(x_3,y_3)$ |
| $(10,0,8,8,8,10,0,8,8)$ | $(2,3,2)$ | $g_2^{(2)}(x_1,y_1)g_3^{(2)}(x_2,y_2)g_2^{(2)}(x_3,y_3)$ |
| $(11,0,8,10,12,11,0,8,10)$ | $(2,3,2)$ | $g_2^{(2)}(x_1,y_1)g_3^{(2)}(x_2,y_2)g_2^{(2)}(x_3,y_3)$ |
| $(13,0,8,2,4,13,0,8,2)$ | $(2,3,2)$ | $g_2^{(2)}(x_1,y_1)g_3^{(2)}(x_2,y_2)g_2^{(2)}(x_3,y_3)$ |
| $(15,0,8,2,12,15,0,8,2)$ | $(2,3,2)$ | $g_2^{(2)}(x_1,y_1)g_3^{(2)}(x_2,y_2)g_2^{(2)}(x_3,y_3)$ |

$g_0^{(2)}(x,y) = x^2 + 720xy + 3375y^2$, $g_1^{(2)}(x,y) = x^2 + 224xy - 225y^2$,
$g_2^{(2)}(x,y) = x^2 - 16xy + 15y^2$, $g_3^{(2)}(x,y) = x^2 - y^2$

Now we define the $m$-spotty Lee weight, the $m$-spotty Lee distance and the $m$-spotty Lee weight enumerator for a byte error-control code over $R_k$.

**Definition 6.2.** For any $\mathbf{u} = (\mathbf{u}_1, \mathbf{u}_2, \cdots, \mathbf{u}_n) \in R_k^{bn}$, the $m$-spotty Lee weight of $\mathbf{u}$ is defined as $w_{ML}(\mathbf{u}) = \sum_{i=1}^{n} \left\lceil \frac{w_L(\mathbf{u}_i)}{t} \right\rceil$, where $\mathbf{u}_i = (u_{i1}, u_{i2}, \cdots, u_{ib}) \in R_k^b$ is the $i$-th byte of $\mathbf{u}$ and $w_L(\mathbf{u}_i) = \sum_{j=1}^{b} w_L(u_{ij})$.

**Definition 6.3.** Let $\mathbf{u}, \mathbf{v}$ be vectors in $R_k^{bn}$ with their $i$-th bytes as $\mathbf{u}_i, \mathbf{v}_i$ respectively. Then the $m$-spotty Lee distance between $\mathbf{u}$ and $\mathbf{v}$, denoted by $d_{ML}$, is defined as

$$d_{ML} = \sum_{i=1}^{n} \left\lceil \frac{d_L(\mathbf{u}_i, \mathbf{v}_i)}{t} \right\rceil = \left\lceil \frac{w_L(\mathbf{u}_i - \mathbf{v}_i)}{t} \right\rceil = w_{ML}(\mathbf{u} - \mathbf{v}).$$

Note that $d_{ML}$ is a metric on $R_k^{bn}$. Further, if $C$ is a byte error-control code of length $bn$ and byte length $b$ over $R_k$, then the number $d_{ML}(C) = \min\{d_{ML}(\mathbf{u}, \mathbf{v}) : \mathbf{u}, \mathbf{v} \in C, \mathbf{u} \neq \mathbf{v}\}$ is called the $m$-spotty Lee distance of the code $C$. Moreover, $d_{ML}(C) = \min\{w_{ML}(\mathbf{u}) : \mathbf{u} \in C, \mathbf{u} \neq 0\}$.

**Definition 6.4.** Let $C$ be a byte error-control code of length $bn$ and byte length $b$ over $R_k$. Then the $m$-spotty Lee weight enumerator of $C$ is defined as

$$L_C(z) = \sum_{\mathbf{u} \in C} z^{w_{ML}(\mathbf{u})} = \sum_{\mathbf{u}=(\mathbf{u}_1, \mathbf{u}_2, \cdots, \mathbf{u}_n) \in C} \prod_{j=1}^{n} z^{\lceil w_L(\mathbf{u}_i)/t \rceil}.$$

Let $\mathbf{u}$ be any vector in $R_k^{bn}$ with $\mathbf{u}_i \in R_k^b$ as its $i$-th byte for $1 \leq i \leq n$. For each $i$, if $j_{ip}(0 \leq p \leq \ell-1)$ is the number of bits in $\mathbf{u}_i$ which are equal to $r_p$, then the $\ell$-tuple $\mathbf{J}_i = (j_{i0}, j_{i1}, \cdots, j_{i,\ell-1})$ is called composition of the $i$-th byte $\mathbf{u}_i$ of $\mathbf{u}$, and the vector $\mathbf{J} = (\mathbf{J}_1, \mathbf{J}_2, \cdots, \mathbf{J}_n)$ is called the composition vector of $\mathbf{u} = (\mathbf{u}_1, \mathbf{u}_2, \cdots, \mathbf{u}_n)$.

Now let $A(\mathbf{J})$ be the number of codewords in $C$ having the composition vector as $\mathbf{J}$. Then the $m$-spotty Lee weight enumerator of $C$ can be rewritten as

$$L_C(z) = \sum_{\mathbf{J}} A(\mathbf{J}) \prod_{j=1}^{n} z^{\lceil \rho(\mathbf{J}_i)/t \rceil},$$

where for each $i$, $\mathbf{J}_i = (j_{i0}, j_{i1}, \cdots, j_{i,\ell-1})$ and $\rho(\mathbf{J}_i) = \sum_{p=0}^{\ell-1} w_L(r_p) j_{ip}$ with $0 \leq j_{ip} \leq b$ for $0 \leq p \leq \ell - 1$.

Let $\mathbf{u}$ be the fixed vector. Define the numbers $s_{pq}, 0 \leq p, q \leq \ell-1$, as the number of components of the vectors $\mathbf{u} = (u_1, u_2, \cdots, u_b)$ and $\mathbf{v} = (v_1, v_2, \cdots, v_b)$ satisfying $u_i = r_p$ and $v_i = r_q$.

**Definition 6.5.** For a fixed positive integer $t$ and $\mathbf{J} = (j_0, j_1, \cdots, j_{\ell-1})$, we define the polynomial

$$g_{\mathbf{J}}^{(t)}(z) = \sum_{s_{pq}} \left( \prod_{p=0}^{\ell-1} \frac{j_p!}{\prod_{q=0}^{\ell-1} s_{pq}!} \chi\left( \sum_{q=0}^{\ell-1} r_p r_q s_{pq} \right) \right) z^{\left\lceil \sum_{p=0}^{\ell-1} \left( \sum_{q=1}^{\ell-1} w_L(r_q) s_{pq} \right) / t \right\rceil},$$

where the summation $\sum_{s_{pq}}$ runs over all non-negative integers $s_{pq}(0 \leq p, q \leq \ell - 1)$ satisfying $\sum_{q=0}^{\ell-1} s_{pq} = j_p$ for every $p$.

**Lemma 6.6.** For a fixed vector $\mathbf{u} \in R_k^b$ with composition $\mathbf{J} = (j_0, j_1, \cdots, j_{\ell-1})$, we have

$$\sum_{\mathbf{v} \in R_k^b} \chi(\langle \mathbf{u}, \mathbf{v} \rangle) z^{\lceil w_L(\mathbf{v}) \rceil} = g_{\mathbf{J}}^{(t)}(z).$$

*Proof.* The proof is similar to that of Lemma 2 in [9]. □

In the following theorem, we derive a MacWilliams type identity for the $m$-spotty Lee weight enumerator of a byte error-control code over an infinite family of rings.

**Theorem 6.7**. Let $C$ be a byte error-control code of length $bn$ over $R_k$ with byte length $b$ and $m$-spotty Lee weight enumerator $L_C(z)$ as defined above. If $A(\boldsymbol{J}_1, \boldsymbol{J}_2, \cdots, \boldsymbol{J}_n)$ denotes the number of codewords in $C$ having the composition vector as $\boldsymbol{J} = (\boldsymbol{J}_1, \boldsymbol{J}_2, \cdots, \boldsymbol{J}_n)$, then the $m$-spotty Lee weight enumerator $L_{C^\perp}$ of the dual code $C^\perp$ is given by

$$L_{C^\perp}(z) = \frac{1}{|C|} \sum_{\boldsymbol{J}} A(\boldsymbol{J}) \prod_{i=1}^{n} g_{\boldsymbol{J}_i}^{(t)}(z),$$

where the summation runs over all $n$-tuples $\boldsymbol{J}$ with each $\boldsymbol{J}_i$, an $\ell$-tuple over $\{0, 1, 2, \cdots, b\}$.

*Proof.* The proof is similar to those of Theorem 3.12 and Theorem 28 in [9]. □

**Example 6.8** Let $C$ be the byte error-control code as defined in Example 4.11. Since $C^\perp = 2147483648$ is very large, we will apply Theorem 6.7 to obtain the $m$-spotty Lee weight enumerator of $C^\perp$. For this, we compute the composition vectors for the codewords of $C$. It is easy to check that $(0, 0, uv, uv, 0, 0, 0, uv, uv) \in C$ and its composition vector is $\boldsymbol{J}_1, \boldsymbol{J}_2$ and $\boldsymbol{J}_3$ where $(\boldsymbol{J}_1, \boldsymbol{J}_2, \boldsymbol{J}_3) = ((*[2]_0*[1]_8*), (*[2]_0*[1]_8*), (*[1]_0*[2]_8*))$. Here, $[s]_p$ represents there are $s$ elements equal to $r_p$, where $0 \leq p \leq 15, 0 \leq s \leq 3$ and $r_p \in R_2$ is defined in Table V, which gives an alternative expression form for every element $r_p = auv + bv + cu + d \in R_2$, where $(a, b, c, d) \in \mathbb{Z}_2^4$.

For $t = 2$, by Definition 6.5, we get

$$g_{\boldsymbol{J}_1}^{(2)}(z) = g_{\boldsymbol{J}_2}^{(2)}(z) = 1 + 6z - 29z^2 + 36z^3 - 9z^4 - 10z^5 + 5z^6 \text{ and}$$
$$g_{\boldsymbol{J}_3}^{(2)}(z) = 1 - 2z - 5z^2 + 20z^3 - 25z^4 + 14z^5 - 3z^6.$$

We note that the codeword $(0, 0, uv, uv, 0, 0, 0, uv, uv) \in C$ contributes the term

$$\begin{aligned} g_{\boldsymbol{J}_1}^{(2)}(z) g_{\boldsymbol{J}_2}^{(2)}(z) g_{\boldsymbol{J}_3}^{(2)}(z) &= 1 + 10z - 51z^2 - 272z^3 + 2132z^4 - 4072z^5 - 4940z^6 + 39312z^7 - 88946z^8 \\ &\quad + 110396z^9 - 74074z^{10} + 9360z^{11} + 28964z^{12} - 25832z^{13} + 7972z^{14} \\ &\quad + 1520z^{15} - 2055z^{16} + 650z^{17} - 75z^{18} \end{aligned}$$

to the $m$-spotty Lee weight enumerator of $C$. Working similarly, we obtain the composition vectors and the contributing polynomials for other codewords in $C^\perp$, which are given in Table VI.

Therefore by Theorem 6.7, the $m$-spotty Lee weight enumerator of the code $C$ is given by

$$\begin{aligned} L_{C^\perp} &= \frac{1}{|C|} \sum_{\boldsymbol{J}} A(\boldsymbol{J}) \prod_{i=1}^{n} g_{\boldsymbol{J}_i}^{(t)}(z) \\ &= 101z^{18} + 5326z^{17} + 122705z^{16} + 1641752z^{15} + 13077404z^{14} + 63554224z^{13} + 196381596z^{12} \\ &\quad + 398386136z^{11} + 538692126z^{10} + 487268316z^9 + 294389014z^8 + 117912840z^7 + 30602524z^6 \\ &\quad + 4946304z^5 + 475132z^4 + 26888z^3 + 1221z^2 + 38z + 1. \end{aligned}$$

# 7 Conclusion

This paper mainly presents the MacWilliams type identities for the $m$-spotty Hamming weight enumerator, joint $m$-spotty Hamming weight enumerator and split $m$-spotty Hamming weight enumerator for byte error-control codes over finite commutative Frobenius rings. Finally, the $m$-spotty Lee weight enumerator over an infinite family of rings $R_k$ is also obtained. In fact, Joint $m$-spotty Lee weight enumerator (Theorem 8 and Theorem 9 in [9]) and Split $m$-spotty Lee weight enumerator (Theorem 5 in [9]) are also valid over the infinite family of rings $R_k$. Moreover, the results in [10] for $r$-fold joint $m$-spotty Hamming weight enumerators still hold over finite commutative Frobenius rings by applying the similar method in Theorem 3.12.

Table V Alternative expression of elements of $R_k$

| $(a, b, c, d)$ | $r_p$ |
|---|---|
| $(0, 0, 0, 0)$ | $r_0 = 0$ |
| $(0, 0, 0, 1)$ | $r_1 = 1$ |
| $(0, 0, 1, 0)$ | $r_2 = u$ |
| $(0, 0, 1, 1)$ | $r_3 = 1 + u$ |
| $(0, 1, 0, 0)$ | $r_4 = v$ |
| $(0, 1, 0, 1)$ | $r_5 = 1 + v$ |
| $(0, 1, 1, 0)$ | $r_6 = u + v$ |
| $(0, 1, 1, 1)$ | $r_7 = 1 + u + v$ |
| $(1, 0, 0, 0)$ | $r_8 = uv$ |
| $(1, 0, 0, 1)$ | $r_9 = 1 + uv$ |
| $(1, 0, 1, 0)$ | $r_{10} = u + uv$ |
| $(1, 0, 1, 1)$ | $r_{11} = 1 + u + uv$ |
| $(1, 1, 0, 0)$ | $r_{12} = v + uv$ |
| $(1, 1, 0, 1)$ | $r_{13} = 1 + v + uv$ |
| $(1, 1, 1, 0)$ | $r_{14} = u + v + uv$ |
| $(1, 1, 1, 1)$ | $r_{15} = 1 + u + v + uv$ |

Table VI Codewords, composition vectors and contributing polynomials

| codewords | $(\boldsymbol{J}_1, \boldsymbol{J}_2, \boldsymbol{J}_3)$ | $g^{(2)}_{\boldsymbol{J}_1} g^{(2)}_{\boldsymbol{J}_2} g^{(2)}_{\boldsymbol{J}_3}$ |
|---|---|---|
| $(0, 0, 0, 0, 0, 0, 0, 0, 0)$ | $(([3]_0*), ([3]_0*), ([3]_0*))$ | $a^3$ |
| $(8, 0, 0, 0, 0, 8, 0, 0, 0)$ | $(([2]_0 * [1]_8*), ([2]_0 * [1]_8*), ([3]_0*))$ | $ab^2$ |
| $(0, 0, 8, 8, 0, 0, 0, 8, 8)$ | $(([2]_0 * [1]_8*), ([2]_0 * [1]_8*), ([1]_0 * [2]_8*))$ | $b^2c$ |
| $(2, 0, 0, 0, 8, 2, 0, 0, 0)$ | $(([2]_0 * [1]_2*), (*[3]_5*), ([1]_0[2]_5*))$ | $ade$ |
| $(10, 0, 0, 0, 8, 10, 0, 0, 0)$ | $(([2]_0 * [1]_{10}*), ([1]_0 * [1]_8 * [1]_{10}*), (*[3]_2*), ([3]_0*))$ | $ade$ |
| $(4, 0, 0, 8, 0, 4, 0, 0, 8)$ | $(([2]_0 * [1]_4*), ([1]_0 * [1]_4 * [1]_8*), ([2]_0 * [1]_8*))$ | $bde$ |
| $(12, 0, 0, 8, 0, 12, 0, 0, 8)$ | $(([2]_0 * [1]_{12}*), ([1]_0 * [1]_8 * [1]_{10}*), ([2]_0 * [1]_8*))$ | $ade$ |
| $(1, 0, 0, 2, 4, 1, 0, 0, 2)$ | $(([2]_0[1]_1*), (*[1]_1[1]_2 * [1]_4*), ([2]_0 * [1]_2*))$ | $cdf$ |
| $(3, 0, 0, 2, 12, 3, 0, 0, 2)$ | $(([2]_0 * [1]_3*), (*[1]_2[1]_3 * [1]_{12}*), ([2]_0 * [1]_2*))$ | $cdf$ |
| $(5, 0, 0, 10, 4, 5, 0, 0, 10)$ | $(([2]_0 * [1]_5*), (*[1]_4[1]_5 * [1]_{10}*), ([2]_0 * [1]_{10}*))$ | $cdf$ |
| $(6, 0, 0, 8, 8, 6, 0, 0, 8)$ | $(([2]_0 * [1]_8*), ([1]_6 * [2]_8*), ([2]_0 * [1]_8*))$ | $bdg$ |
| $(7, 0, 0, 10, 12, 7, 0, 0, 10)$ | $(([2]_0 * [1]_7*), (*[1]_7 * [1]_{10} * [1]_{12}*), ([2]_0 * [1]_{10}*))$ | $deg$ |
| $(9, 0, 0, 2, 4, 9, 0, 0, 2)$ | $(([2]_0 * [1]_9*), (*[1]_2 * [1]_4 * [1]_9*), ([2]_0 * [1]_2*))$ | $deg$ |
| $(11, 0, 0, 2, 12, 11, 0, 0, 2)$ | $(([2]_0 * [1]_{11}*), (*[1]_2 * [1]_{11} * [1]_{12}*), ([2]_0 * [1]_2*))$ | $deg$ |
| $(13, 0, 0, 10, 4, 13, 0, 0, 10)$ | $(([2]_0 * [1]_{13}*), (*[1]_4 * [1]_{10} * [1]_{13}*), ([2]_0 * [1]_{10}*))$ | $deg$ |
| $(14, 0, 0, 8, 8, 14, 0, 0, 8)$ | $(([2]_0 * [1]_{14}*), (*[2]_8 * [1]_{14}*), ([2]_0 * [1]_8*))$ | $bdg$ |
| $(15, 0, 0, 10, 12, 15, 0, 0, 10)$ | $(([2]_0 * [1]_{15}*), (*[1]_{10} * [1]_{12} * [1]_{15}*), ([2]_0 * [1]_{10}*))$ | $dcf$ |
| $(4, 0, 8, 0, 0, 4, 0, 8, 0)$ | $(([1]_0 * [1]_4 * [1]_8*), ([2]_0 * [1]_4*), ([2]_0 * [1]_8))$ | $bde$ |
| $(12, 0, 8, 0, 0, 12, 0, 8, 0)$ | $(([1]_0 * [1]_8 * [1]_{12}*), ([2]_0 * [1]_{12}*), ([2]_0 * [1]_8))$ | $bde$ |
| $(6, 0, 8, 0, 8, 6, 0, 8, 0)$ | $(([1]_0 * [1]_6 * [1]_8*), ([1]_0 * [1]_6 * [1]_8*), ([2]_0 * [1]_8))$ | $be^2$ |
| $(14, 0, 8, 0, 8, 14, 0, 8, 0)$ | $(([1]_0 * [1]_8 * [1]_{14}*), ([1]_0 * [1]_8 * [1]_{14}*), ([2]_0 * [1]_8))$ | $be^2$ |
| $(8, 0, 8, 8, 0, 8, 0, 8, 8)$ | $(([1]_0 * [2]_8*), ([1]_0 * [2]_8*), ([1]_0 * [2]_8*))$ | $c^3$ |
| $(1, 0, 8, 10, 4, 1, 0, 8, 10)$ | $(([1]_0[1]_1 * [1]_8*), (*[1]_1 * [1]_4 * [1]_{10}*), ([1]_0 * [1]_8 * [1]_{10}*))$ | $c^2e$ |
| $(2, 0, 8, 8, 8, 2, 0, 8, 8)$ | $(([1]_0 * [1]_2 * [1]_8*), (*[1]_2 * [2]_8*), ([1]_0 * [2]_8*))$ | $ceg$ |
| $(3, 0, 8, 10, 12, 3, 0, 8, 10)$ | $(([1]_0 * [1]_3 * [1]_8*), (*[1]_3 * [1]_{10} * [1]_{12}*), ([1]_0 * [1]_8 * [1]_{10}*))$ | $c^2e$ |
| $(5, 0, 8, 2, 4, 5, 0, 8, 2)$ | $(([1]_0 * [1]_5 * [1]_8*), (*[1]_2 * [1]_4 * [1]_5*), ([1]_0 * [1]_2 * [1]_8*))$ | $c^2e$ |
| $(7, 0, 8, 2, 12, 7, 0, 8, 2)$ | $(([1]_0 * [1]_7 * [1]_8*), (*[1]_2 * [1]_7 * [1]_{12}*), ([1]_0 * [1]_2 * [1]_8*))$ | $e^3$ |
| $(9, 0, 8, 10, 4, 9, 0, 8, 10)$ | $(([1]_0 * [1]_8[1]_9*), (*[1]_4 * [1]_9 * [1]_{10}*), ([1]_0 * [1]_8 * [1]_{10}*))$ | $e^3$ |
| $(10, 0, 8, 8, 8, 10, 0, 8, 8)$ | $(([1]_0 * [1]_8 * [1]_{10}*), (*[2]_8 * [1]_{10}*), ([1]_0 * [2]_8*))$ | $ceg$ |
| $(11, 0, 8, 10, 12, 11, 0, 8, 10)$ | $(([1]_0 * [1]_8 * [1]_{11}*), (*[1]_{10}[1]_{11}[1]_{12}*)), ([1]_0 * [1]_8 * [1]_{10}*))$ | $e^3$ |
| $(13, 0, 8, 2, 4, 13, 0, 8, 2)$ | $(([1]_0 * [1]_8 * [1]_{13}*), (*[1]_2 * [1]_4 * [1]_{13}*)), ([1]_0 * [1]_2 * [1]_8*))$ | $e^3$ |
| $(15, 0, 8, 2, 12, 15, 0, 8, 2)$ | $(([1]_0 * [1]_8 * [1]_{15}*), (*[1]_2 * [1]_{12} * [1]_{15}*)), ([1]_0 * [1]_2 * [1]_8*))$ | $c^2e$ |

$a = 1 + 78z + 715z^2 + 1716z^3 + 1287z^4 + 286z^5 + 13z^6$, $b = 1 + 6z - 29z^2 + 36z^3 - 9z^4 - 10z^5 + 5z^6$,
$c = 1 - 2z - 5z^2 + 20z^3 - 25z^4 + 14z^5 - 3z^6$, $d = 1 + 34z + 55z^2 - 132z^3 - 33z^4 + 66z^5 + 9z^6$,
$e = 1 - 6z + 15z^2 - 20z^3 + 15z^4 - 6z^5 + z^6$, $f = 1 + 54z + 275z^2 + 132z^3 - 297z^4 - 154z^5 - 11z^6$.

## 8  Acknowledgments

This research was done while the author was visiting CCRG of Nanyang Technological University. The author is grateful to Professor San Ling for helpful discussions which improved the presentation of the material.

# References

[1] S. T. Dougherty, B. Yildiz, and S. Karadeniz, Codes over $R_k$, Gray maps and their binary images. Designs, Codes and Cryptography April 2012, 63(1):113-126.

[2] S. T. Dougherty, M. Harada, M. Oura, Note on the $g$-fold joint weight enumerators of self-dual codes over $\mathbb{Z}_k$, Appl. Algebra Eng. Commun. Comput., 2001, 11(6):437-445.

[3] Y. Fan, S. Ling, and H. W. Liu, Matrix product codes over finite commutative Frobenius rings, Designs, Codes and Cryptography, Doi 10.1007/s10623-012-9726-y.

[4] E. Fujiwara, Code design for dependable system, Theory and practocal application, Wiley & Son, Inc., 2006.

[5] F. J. MacWilliams, N. J. Sloane, The Theory of error-correcting codes, North-Holland Publishing Company, Amsterdam, 1978.

[6] M. Özen, V. Siap, The MacWilliams identity for $m$-spotty weight enumerators of linear codes over finite fields, Computers and Mathematics with Applications, 2011, 61(4): 1000-1004.

[7] M. Özen, V. Siap, The MacWilliams identity for $m$-spotty Rosenbloom-Tsfasman weight enumerator, Journal of the Franklin Institute, (2012), http://dx. doi.org/10. 1016/j. jfranklin. 2012. 06. 002.

[8] A. Sharma and A. K. Sharma, MacWilliams type identities for some new $m$-spotty weight enumerators, IEEE Transactions on Information Theory, 2012, 58(6): 3912-3924.

[9] A. Sharma and A. K. Sharma, On some new $m$-spotty Lee weight enumerators, Designs, Codes and Cryptography, Doi 10. 1007/s10623-012-9725-z.

[10] A. Sharma and A. K. Sharma, On MacWilliams type identities for $r$-fold joint $m$-spotty weight enumerators, Discrete Mathematics, 2012, 312(22): 3316-3327.

[11] M. J. Shi, MacWilliams identity for $m$-spotty RT weight enumerators over finite commutative Frobenius rings, submitted to Science China Mathematics for publication.

[12] M. J. Shi, MacWilliams identity for $m$-spotty weight enumerators over $\mathbb{F}_2+u\mathbb{F}_2+\cdots+u^{m-1}\mathbb{F}_2$, submitted to Turkish Journal of Mathematics for publication.

[13] I. Siap, An identity between the $m$-spotty weight enumerators of a linear code and its dual, Turkish Journal of Mathematics, doi:10.3906/mat-1103-55.

[14] I. Siap, MacWilliams identity for $m$-spotty Lee weight enumerators, Applied Mathematics Letters, 2010, 23(1): 13-16.

[15] K. Suzuki, T. Kashiyama, E. Fujiwara, MacWilliams identity for $m$-spotty weight enumerator. in: ISIT 2007. Nice, France, 2007, pp:31-35.

[16] K. Suzuki, T. Kashiyama, E. Fujiwara, A general class of $m$-spotty weight enumerator, IEICE-Transactionson Fundamentals of Electronics, Communications and Computer Sciences E90-A(7)(2007): 1418-1427.

[17] J. Wood, Duality for modules over finite rings and application to coding theory. Am. J. Math. 1999, 121. 551-575.

[18] J. Wood, Application of finite Frobenius rings to the foundations of algebraic coding theory. Proceedings of the 44th Symposium on Ring Theory and Representation Theory (Okayama University, Japan, September 25-27, 2011), O. Iyama, ed., Nagoya, Japan, 2012, pp. 223-245.